\documentclass[11pt]{article}
\usepackage{bm}
\usepackage{amsmath}
\usepackage{subcaption}
\usepackage{setspace}
\usepackage{authblk}
\usepackage[margin=1.0in]{geometry}

\usepackage{empheq}
\usepackage{hyperref, comment, tikz}
\usepackage[dvipsnames]{xcolor}

\newcommand{\gd}{\dot{\gamma}}
\newcommand{\bZ}{\boldsymbol{\mathbf{0}}}
\newcommand{\btaun}{\boldsymbol{\mathbf{\tau}}}

\newcommand{\bun}{\boldsymbol{u}}

\begin{document}

\title{A Modified Suspension Balance Model for Deformable Particle Suspensions: Application to Blood Flows with Cell-Free Layer}

\author[1]{Hugo A Castillo-S{\'a}nchez}
\author[3]{Weston Ortiz}
\author[3]{Richard Martin}
\author[1]{Rukiye Tuna}
\author[4]{Rekha R Rao}
\author[1,2]{Z Leonardo Liu\thanks{Corresponding author. Email: \texttt{leo.liu@eng.famu.fsu.edu}}}

\affil[1]{Department of Chemical and Biomedical Engineering, FAMU--FSU College of Engineering, Tallahassee, FL, USA}
\affil[2]{Institute for Successful Longevity, Florida State University, Tallahassee, FL, USA}
\affil[3]{Center for MicroEngineered Materials, University of New Mexico, Albuquerque, NM, USA}
\affil[4]{Sandia National Laboratories, Albuquerque, NM, USA}

\maketitle

\begin{abstract}
We propose a modified suspension balance model (SBM) for the flow of red blood cells (RBCs) and other deformable particle suspensions in confined geometries. Specifically, the method includes the hydrodynamic lift force generated by deformable particles interacting with walls leading to a cell-free layer. The lift force is added to the SBM to drive RBCs migrating away from the wall. Using the modified SBM (MSBM), we simulate blood flows through microvascular channels and tubes. The method is able to capture the transient development of the cell-free layer (CFL) and the corresponding hematocrit and velocity profiles with the development of the CFL. The CFL thickness and hemorheological hallmarks in microcirculation, such as the Fåhræus Effect and the Fåhræus-Linqvist Effect, are captured in good agreement with existing experimental and direct numerical results of blood flows. This work establishes a novel continuum computational framework that can efficiently capture the microstructural heterogeneity and non-Newtonian flow behavior of concentrated deformable particle suspensions under confinement.

\vspace{0.1cm}
\end{abstract}

\section{Introduction}
Suspension flows of non-Brownian or non-colloidal particles have incited much interest in the past decades. 
Suspensions arise in many applications, including blood flow, food processing, pharmaceuticals, and detergents~\cite{rao1977,goldsmith1968,kulshreshtha2010}. In non-colloidal suspension flows, the radii of the suspended particles are typically
larger than one micron and the effect of thermal fluctuation is negligible. Even without colloidal effects, these flows can be difficult to model for a number of reasons including inherent nonlinearity, multi-body interactions, particle polydispersity, fluid-particle interactions, and non-Newtonian suspending fluids~\cite{mueller2010,stickel2005,gillissen2018}. 

In concentrated particle suspensions, one unique rheological aspect lies in the presence of particle stress in addition to the fluid stress generated by the disturbance and interactions of suspended particles, as elegantly conceptualized and deduced by Batchelor~\cite{batchelor}. The presence of particle stress can lead to particle cross-stream migration, where particles in suspension irreversibly migrate across the streamlines to preferred locations determined by gradients in viscosity, concentration, and shear rate~\cite{phillips1992,Leigh1987a, Leigh1987b}. For non-dilute particle suspension flow through a channel, particles at the high-shear regions generally tend to migrate to low-shear-rate regions, while particle inertia~\cite{segre1962behaviour},  particle-wall interaction~\cite{liu2019}, particle rigidity~\cite{kumar2012ms} and other effects~\cite{tahiri2013,audemar2022,gou2023}, all play important roles to alter the particle migration dynamics.


Blood is a biological non-colloidal suspension of deformable cells, and the interactions between the suspended red blood cells can give rise to shear-thinning behavior as well as other interesting rheological phenomena, including thixotropy, viscoelasticity and non-zero yield stress~\cite{dintenfass1962,thurston1972,merrill1963,horner2018}. It is well known that blood flow in microvessels exhibits a cell-free layer (CFL), as reported from previous experimental works~\cite{aarts,noguchi2023,merlo2023} where RBCs migrate away from the wall and deplete the RBC concentration (hematocrit) near the wall~\cite{secomb2017blood}. The formation of CFL is physiologically essential to maintain a relatively low apparent viscosity~\cite{Pries2005} and to promote platelet and leukocyte margination needed for healthy hemostasis~\cite{aarts,kumar2012,kumar2012ms,liu2022sipa}.
Mechanistically, RBC migration against the wall is driven by a deformability-induced hydrodynamic lift force arising from asymmetric particle-wall interactions~\cite{Abkarian2002}. During this process, the cell deformability and the RBC suspension dynamics (such tank treading motion) play important roles in generating and controlling the lift force and consequently the CFL formation~\cite{reasor2013determination}.

To properly model the hemodynamics and hemorheology in microcirculation, the lift force-induced RBC migration needs to be considered to capture the CFL. Cellular-scale direct computational models have been able to directly capture the RBC migration, forming physiologically consistent CFL phenomenon~\cite{fedosov2010,henriquez2016,liu2018nanoparticle,balogh2019}. However, direct computational models can be computationally prohibitive for vascular network or organ-scale simulations. Continuum scale models provide an alternative that can be feasibly extended to at-scale applications~\cite{qohar2021,qi2017theory,qi2018time,karmakar2024}. Qin and Shaqfeh~\cite{qi2018time} presented a continuum model to simulate the time evolution of RBCs and platelet concentration distributions in pressure-driven flow. Their model considers deformability-induced lift velocity and shear-induced diffusion as governing mechanisms for the cross-flow transport of RBCs. 
Karmakar {\it et. al.}~\cite{karmakar2024} presented a homogenized two-phase computational framework for blood flow simulations to predict RBC velocity, spatial distribution, and cell-free layer thickness, where a wall-induced lift force expression for RBCs as a function of hematocrit is introduced. 


Rooted in the rigorous particle stress framework established by Batchelor~\cite{batchelor}, the suspension balance model (SBM) has been successfully developed and applied to model a variety of flows for rigid particle suspension, including channel flows \cite{Miller2006,Morris1998}, pipe flows \cite{Miller2006}, and contraction flows \cite{Miller2006, Miller2009}. Morris and Boulay \cite{Morris1999} modeled curvilinear flows, including flows in a wide-gap Couette, between parallel circular plates, and in a cone-and-plate device. The SBM first proposed by Nott and Brady \cite{Nott1994} postulates that particle migration can be modeled by accounting for the particles stresses induced by the particle-particle hydrodynamic interactions at high concentration. In this work, we propose a modified suspension balance model (MSBM) that incorporates an additional lift forcing term to to the conventional SBM method to account for the wall-RBC hydrodynamics interactions near the wall. 

The paper is organized as follows. A detailed derivation of the SBM model and the MSBM is presented in section~\ref{sec:methods}, where we discuss how the lift force term is developed and incorporated into the conventional SBM. In addition, we briefly discuss the numerical implementation and the numerical solvers we used to simulate blood flows using the MBSM. In section~\ref{sec:2Dchannel}, we present how the effect of various MSBM parameters of biophysical significance affect the hematocrit, velocity distribution and CFL thickness in a channel flow setting. In section~\ref{sec:theoryvsexp}, we directly compare our continuum results with existing theory and experiments in both channel flow and simple shear flow configurations. In section~\ref{sec:tubularflow}, we further demonstrate that our MSBM prediction in a tubular flow configuration can capture the physiological F\aa hr\ae us effect and F\aa hr\ae us-Linqvist effect. Concluding remarks are given in section~\ref{sec:conclusions}.


\section{Methodology} \label{sec:methods}
\subsection{Conventional Suspension Balance Model}
\label{sec:csbm}

The suspension balance model is derived by considering the conservation of mass
and momentum both for the suspension bulk and for the particle phases separately \cite{Nott1994}.
The mass and momentum balance equations for the suspension bulk in terms of averaged quantities can be obtained as
\begin{eqnarray}
    \nabla \cdot \langle \bm{u} \rangle = 0 \label{eq:cont} \\
    \frac{\partial \langle \rho \bm{u} \rangle}{\partial t} + \langle \bm{u} \rangle
    \cdot \nabla \langle \rho \bm{u} \rangle = \langle \bm{b} \rangle + \nabla \cdot \langle \bm{\Sigma} \rangle
    \mbox{ ,} \label{eq:momentum}
\end{eqnarray}
where $\langle \cdot \rangle$ denotes a suspension volume average, $\bm{u}$ is the suspension velocity, 
$\rho$ is the suspension density, and $\bm{b}$ is the body force. The suspension stress,
$\langle \bm{\Sigma} \rangle$, comprises the contributions from both the fluid and particle phases
\begin{equation}
    \langle \bm{\Sigma} \rangle = -\langle p \rangle_{f} \bm{I} + 2\mu_{f} \langle \bm{e} \rangle
    + \langle \bm{\Sigma} \rangle_{p} \mbox{ ,} \label{eq:suspension_str}
\end{equation}
where $\langle p \rangle_{f}$ is the averaged pressure in the fluid phase, 
$\mu_{f}$ is the dynamic viscosity of the suspending fluid, and
$\langle \bm{e} \rangle = \left( \nabla \langle \bm{u} \rangle + \nabla \langle
\bm{u} \rangle^{T} \right) / 2$ is the
bulk rate of strain tensor. The stress due to the presence of the particle phase, $\langle \bm{\Sigma} \rangle_{p}$,
will be discussed below. Note that $\langle \cdot \rangle_{f}$ is a fluid-phase average and 
$\langle \cdot \rangle_{p}$ is a particle-phase average. 

Since we consider suspensions with neutrally buoyant effects and vanishing inertial effects,
the governing equations (\ref{eq:cont}) and (\ref{eq:momentum}) can be written as
\begin{eqnarray}
    \nabla \cdot \langle \bm{u} \rangle = 0  \mbox{ ,}  \label{eq:cont2} \\
    \nabla \cdot \langle \bm{\Sigma} \rangle = 0 \label{eq:momentum2} \mbox{ .}
\end{eqnarray}

The mass and momentum balance equations for the particle phase in terms of the volume fraction can be obtained as
\begin{eqnarray}
    \frac{\partial \phi}{\partial t} + \nabla \cdot \left( \langle \bm{u} \rangle_{p} \phi \right)= 0 \mbox{ ,}   \\
    \rho_{p} \left[ \phi \frac{\partial \langle \bm{u} \rangle_{p}}{\partial t} +\langle \bm{u} \rangle_{p}
    \cdot \nabla \langle \bm{u} \rangle_{p} \right] = \langle \bm{b} \rangle_{p} + \langle \bm{F}^{H} \rangle_{p}
    \mbox{ ,}
\end{eqnarray}
where $\phi$ is the particle volume concentration, $\rho_{p}$ is the particle density, 
$\langle \bm{u} \rangle_{p}$ is the particle-phase average velocity, and $\langle \bm{b} \rangle_{p}$ is the particle-phase body force that is set to zero in the current study. The term $\langle \bm{F}^{H} \rangle_{p}$ is the total hydrodynamic force that captures the inter-phase momentum transfer. 

Similarly, the particle-phase inertial and body force effects can be assumed to be vanishing, which yields a simplified particle-phase momentum balance equation 
\begin{eqnarray}
    \frac{\partial \phi}{\partial t} + \nabla \cdot \left( \langle \bm{u} \rangle_{p} \phi \right) = 0 \mbox{ ,} \label{eq:particle_momentum1} \\
    \langle \bm{F}^{H} \rangle_{p} = 0 \mbox{ .} \label{eq:hydro_force1}
\end{eqnarray}
Equation (\ref{eq:particle_momentum1}) can be written as
\begin{equation}
	\frac{\partial \phi}{\partial t} + \langle \bm{u} \rangle \cdot \nabla \phi
	= -\nabla \cdot \bm{j}_{\perp} \mbox{ ,} \label{eq:convection_diffusion}
\end{equation}
where $\bm{j}_{\perp} = \phi (\langle \bm{u} \rangle_{p} - \langle \bm{u} \rangle )$ is
the particle migration flux relative to the bulk motion \cite{Miller2006}.
To close this system, a constitutive relationship for the inter-phase hydrodynamic force is needed, which is the key idea of the SBM method. Nott {\it et. al.}~\cite{Nott2011} decompose $\langle \bm{F}^{H} \rangle_{p}$ into two
components, including a sedimentation hindrance drag, $\langle \bm{F}^{d} \rangle_{p}$, and a divergence of a particle-phase stress,
$\nabla \cdot \langle \bm{\Sigma} \rangle_{p}$, that is
\begin{equation}
    \langle \bm{F}^{H} \rangle_{p} \approx \langle \bm{F}^{d} \rangle_{p} + \nabla \cdot \langle \bm{\Sigma}
    \rangle_{p} \mbox{ ,} \label{eq:hydro_force}
\end{equation}
where $a$ is the particle radius.
The term $\langle \bm{F}^{d} \rangle_{p}$ can be linearly related to the difference between $\langle \bm{u} \rangle_{p}$
and $\langle \bm{u} \rangle$ by setting
\begin{equation}
	\langle \bm{F}^{d} \rangle_{p} = -\frac{9 \mu_{f}}{2a^{2}} \phi \left[ f(\phi)\right]^{-1}
	\left[ \langle \bm{u} \rangle_{p} - \langle \bm{u} \rangle \right] \mbox{ .}
	\label{eq:sediment_drag}
\end{equation}
Here, $f(\phi)$ is the hindered settling function in the form of Miller and Morris \cite{Miller2006} given by
\begin{equation}
	f\left( \phi \right) = \left( 1 - \frac{\phi}{\phi_{m}} \right) \left(1- \phi\right)^{\alpha -1} \mbox{ ,}
    \label{eq:hinderedsetf}
\end{equation}
where $\phi_{m}$ is the maximum packing concentration and the parameter $\alpha$ has values in the range $2 \le \alpha \le 4$.
Substituting Eqs.~(\ref{eq:hydro_force}) and (\ref{eq:sediment_drag}) into Eq.~(\ref{eq:hydro_force1}) yields
\begin{equation}
	-\frac{9 \mu_{f}}{2a^{2}} \phi \left[ f(\phi)\right]^{-1}
	\left[ \langle \bm{u} \rangle_{p} - \langle \bm{u} \rangle \right]
	+ \nabla \cdot \langle \bm{\Sigma} \rangle_{p} = 0 \mbox{ ,}
\end{equation}
from which $\bm{j}_{\perp}$ can be evaluated as:
\begin{equation}
	\bm{j}_{\perp} = \frac{2a^{2}}{9\mu_{f}} f(\phi) \left[ \nabla \cdot
	\langle \bm{\Sigma} \rangle_{p} \right] \mbox{ .} \label{eq:j_flux}
\end{equation}

The particle-phase stress can be expressed in terms of a concentration-dependent viscosity 
function, a flow-aligned tensor $\bm{Q}$, and the average rate of strain \cite{Morris1999}, as follows:
\begin{equation}
    \langle \bm{\Sigma} \rangle_{p} = -\mu_{f} \mu_{n}(\phi)  \bm{Q}\big(\gd + \gd_{NL} \big)
    + 2\mu_f \left( \mu_{r}(\phi) - 1 \right) \langle \bm{e} \rangle \mbox{ ,} \label{eq:particle_str}
\end{equation}
where $\mu_{n}(\phi)$ is a normal viscosity given by Morris and Boulay~\cite{Morris1999},
\begin{equation}
    \mu_{n}(\phi) = K_{n} \left( \frac{\phi}{\phi_{m}} \right)^{2}
    \left( 1 - \frac{\phi}{\phi_{m}} \right)^{-2} \mbox{ ,} \label{eq:norm_visc}
\end{equation}
with $K_{n}=0.75$. The flow-aligned tensor $\bm{Q}$ captures the anisotropy of the normal stresses~\cite{Miller2006} with the form:

\begin{equation}
\bm{Q} = \begin{bmatrix}
\lambda_{1} &  0    &  0  \\
0  &  \lambda_{2}  &  0  \\
0  &  0    &  \lambda_{3}
\end{bmatrix}
\mbox{.} \label{eq:qtensor}
\end{equation}

The directions of $\bm{Q}$ correspond to the principal directions of viscometric shear flow, where $1$, $2$ and $3$ denote flow, gradient and vorticity directions, respectively. The diagonal elements of $\bm{Q}$ take the following values: $ \lambda_{1}=1$,  $ \lambda_{2} \approx 0.8$ and $ \lambda_{3} \approx 0.5$, which provide reasonably good agreement with concentrated suspension rheology~\cite{zarraga2000} and with observed migration behavior in viscometric flows~\cite{phillips1992}. The relative viscosity $\mu_{r}(\phi)$ is modeled by Krieger's correlation 
\begin{equation}
	\mu_{r} \left( \phi \right) = \left( 1 - \frac{\phi}{\phi_{m}}  \right)^{-2}  \mbox{ .} \label{eq:krieger}
\end{equation}
The shear rate $\dot{\gamma}$
is given by 
\begin{equation}
\dot{\gamma}=\sqrt{2\langle \bm{e} \rangle: \langle \bm{e} \rangle} \mbox{ .} 
\end{equation}
As stated by Dbouk {\it et al.}~\cite{dbouk}, there might exist some regions where the flow experiences a zero shear rate (for instance, the centre-line of a channel), and in these scenarios, the model will predict concentration profiles with a cusp representing a singularity at $\phi=\phi_m$. This singularity, which is also observed in equivalent models (Qi {\it et al.}~\cite{qi2017theory}), is located in a very narrow zone whose size is of the order of magnitude of the particle size $a$. The simplest way to reduce this singularity in these zero shear-rate regions is to add a non-local shear rate $\gd_{NL}$ to the local shear rate $\gd$ (see ~\cite{miller,qi2017theory}), as shown in equation~(\ref{eq:particle_str}), where $\gd_{NL}$ can be defined as:
\begin{equation}
\gd_{NL}=a_s \, \frac{U_{0}}{H}.
\end{equation}

Here, $U_{0}$ is a characteristic velocity (i.e. the center-line velocity in a channel), $H$ is a characteristic length (i.e. channel height) and $a_s$ is a parameter equal to $0$, $\epsilon$ or $\epsilon^2$, where $\epsilon$ is a dimensionless length that take the form $\epsilon = a/H$ in the present work. 

\subsection{Modified Suspension Balance model including Lift Force}
\label{sec:msbm}
To modify the suspension balance model for suspensions of deformable particles (such as capsules, bubbles, vesicles or RBCs in this work), we add a particle deformability-induced lift force in addition to the Stokesian-type drag force to account for the cell free layer phenonmenon. As a result, equation (\ref{eq:hydro_force}) is augmented as 
\begin{equation}
    \langle \bm{F}^{H}_{A} \rangle_{p} \approx \langle \bm{F}^{d} \rangle_{p} + \nabla \cdot \langle
    \bm{\Sigma} \rangle_{p} + \phi \bm{L}_{\perp} \label{eq:modified_force} \mbox{ ,}
\end{equation}
where the volume-averaged lift forcing term $\bm{L}_{\perp}$ can be approximated based on the direct measurement of the lift force exerted on deformable vesicles/capsules. For example, Abkarian {\it et al.}~\cite{Abkarian2002} studied the behavior of vesicles in a shear flow close to a wall, and their experimental observations allowed them to determine a viscous lift force $F_l$ of the form $F_l=\mu_{f} \, \gd \,(a^3/h) f(1-\nu)$, where $\mu_{f}$ is the external fluid viscosity, $a$ is the radius of the vesicle, $h$ is its distance from the substrate ($h=R-r$), and $f(1-\nu)$ is a monotonically decreasing function of the reduced volume $\nu$.

Other experimental and theoretical works have also studied the phenomenon of non-inertial lateral migration of vesicles~\cite{coupier2008} and red blood cells~\cite{qi2017theory,qi2018time,grandchamp2013,losserand2019} in bounded Poiseuille flow, and they have postulated a lateral migration lift velocity, $u_{lift}$, that takes the following dimensionless form $u_{lift} = \xi \gd (r)/h^{\beta}$. Using a cylindrical coordinate system, where $z$ is the flow direction, and $r$ is the gradient direction, a fluid is moving with velocity $u_z$ in a 2D channel with $L  \times R$ dimensions creating a lift velocity in the r-direction normal to the wall ($L$ being the length and $R$ the radius of the channel). For this case, this law illustrates that the migration of particles away from the wall depends on the local shear rate $\gd (r)$, on the particle-wall distance $y$, and on two fitting parameters $\xi$ and $\beta$. This equation implies that as particles move away from the wall, the migration velocity will decrease to zero as it approaches the centerline.

Inspired by both the lift force model developed in Abkarian et al.~\cite{Abkarian2002} and the lift velocity works by Coupier and others~\cite{coupier2008,qi2017theory,qi2018time,grandchamp2013,losserand2019}, we propose a volume-averaged lift force term $\bm{L}_{\perp}$ with the following form:
\begin{equation}
    \bm{L}_{\perp} = \frac{3\,\mu_f \,\gd}{4 \,\pi \, (h+h_0)^{\beta}} \, f(1-\nu) \, \bm{i}_{\perp} \mbox{ ,} \label{eq:lift_force1}
\end{equation}
where $h$ is the normal distance from the wall, $h_0$ is a geometric factor that accounts for the finite-size effects of RBCs and also numerically avoid division by zero, $f(1-\nu)$ is a function of the reduced volume $\nu$, and $\bm{i}_{\perp}$ is a unit vector normal pointing away from the wall. We introduce a power-law coefficient, $\beta$, to adjust the force-distance dependence to account for particles beyond just simple vesicles (such as RBCs in this work). In the current study, we set $h_0=1\times10^{-12}$ m and the $f(1-\nu)$ is set to 1.2 with a unit of $h^{\beta-1}$. We varies $\beta$=1$\sim$1.2 to reflect particle dependence.

Therefore, we have obtained the modified particle-phase governing equations for the MSBM formulation (\ref{eq:modified_force})
\begin{eqnarray}
	\frac{\partial \phi}{\partial t} + \langle \bm{u} \rangle \cdot \nabla \phi
	= -\nabla \cdot \bm{j}_{\perp}^{\, \prime} \mbox{ ,}\\
	\langle \bm{F}^{d} \rangle_{p} + \nabla \cdot \langle \bm{\Sigma} \rangle_{p} + \phi \bm{L}_{\perp} = 0 \mbox{ ,} \label{eq:lift_force}
\end{eqnarray}
where $\bm{j}_{\perp}^{\, \prime}$ is the modified particle migration flux for RBCs.
Substituting Eq.~(\ref{eq:sediment_drag}) into Eq.~(\ref{eq:lift_force}) results in
\begin{equation}
	-\frac{9 \mu_{f}}{2a^{2}} \phi \left[ f(\phi)\right]^{-1}
	\left[ \langle \bm{u} \rangle_{p} - \langle \bm{u} \rangle \right]
	+ \nabla \cdot \langle \bm{\Sigma} \rangle_{p} + \phi \bm{L}_{\perp} = 0 \mbox{ .}
\end{equation}
The modified particle migration flux can then be evaluated as
\begin{equation}
    	\bm{j}_{\perp}^{\, \prime} = \frac{2a^{2}}{9\mu_{f}} f(\phi) \left[ \nabla \cdot
    \langle \bm{\Sigma} \rangle_{p} + \phi \bm{L}_{\perp} \right] \mbox{ .} \label{eq:j2}
\end{equation}
where $\langle \bm{\Sigma} \rangle_{p}$ and $\bm{L}_{\perp}$
are determined by Eqns.~(\ref{eq:particle_str}) and 
(\ref{eq:lift_force1}), respectively. As a first-order approximation, we follow Krieger's model for the relative viscosity, while deformable particle viscosity models can be utilized as future improvement. In summary, our proposed MSBM therefore solves a system of equations with defined constitutive equations as follows,

\begin{equation}
    \nabla \cdot \langle \bm{u} \rangle = 0  \mbox{ ,}   \qquad
    \nabla \cdot \langle \bm{\Sigma} \rangle = 0 \mbox{,}
    \label{eq:constmodelmsbm1}
\end{equation}

\begin{equation}
    \langle \bm{\Sigma} \rangle = -\langle p \rangle_{f} \bm{I} + 2\mu_{f} \langle \bm{e} \rangle
    + \langle \bm{\Sigma} \rangle_{p} \mbox{ ,}
\end{equation}

\begin{equation}
    \langle \bm{\Sigma} \rangle_{p} = -\mu_{f} \mu_{n}(\phi)  \bm{Q}\big(\gd + \gd_{NL} \big)
    + 2\mu_f \left( \mu_{r}(\phi) - 1 \right) \langle \bm{e} \rangle \mbox{ ,}
\end{equation}

\begin{equation}
\bm{Q} = \begin{bmatrix}
1 &  0    &  0  \\
0  &  0.8  &  0  \\
0  &  0    &  0.5
\end{bmatrix}
\end{equation}

\begin{equation}
    \mu_{n}(\phi) = K_{n} \left( \frac{\phi}{\phi_{m}} \right)^{2}
    \left( 1 - \frac{\phi}{\phi_{m}} \right)^{-2} \qquad \mu_{r} \left( \phi \right) = \left( 1 - \frac{\phi}{\phi_{m}}  \right)^{-2}  \mbox{ .} 
\end{equation}

\begin{equation}
	\frac{\partial \phi}{\partial t} + \langle \bm{u} \rangle \cdot \nabla \phi
	= -\nabla \cdot \bm{j}_{\perp}^{\, \prime} \mbox{ ,}\qquad \bm{j}_{\perp}^{\, \prime} = \frac{2a^{2}}{9\mu_{f}} f(\phi) \left[ \nabla \cdot
    \langle \bm{\Sigma} \rangle_{p} + \phi \bm{L}_{\perp} \right] \mbox{ .}    
\end{equation}

\begin{equation}
    \bm{L}_{\perp} = \frac{3\,\mu_f \,\gd}{4 \,\pi \, (h+h_0)^{\beta}} \, f(1-\nu) \, \bm{i}_{\perp} \mbox{ ,}
    \label{eq:constmodelmsbm2}
\end{equation}



\subsection{Numerical Implementation}
\label{sec:OF}
The proposed MSBM method can be conveniently implemented into various multi-phase Computational Fluid Dynamics (CFD) software packages~\cite{Gomasystem,GOMAuserguide,osti_1639568,jasak1996}. 
Specifically for this work, we implemented the MSBM into {\it OpenFOAM}, which is a leading open-source package based on classical Finite-Volume-Method (FVM) approach~\cite{jasak1996}. It is created for ease of developing customized numerical solvers for the solution of continuum mechanics problems. Specifically, we extend the {\it SbmFoam} solver~\cite{dbouk} to develop the {\it SbmBloodFoam} solver, which aims to become a continuum solver for simulating blood flows with cell-level hemorheological details. Our solvers are open-source and can be found in the following link: \url{https://github.com/livingfluids/SbmBloodFoam}.


\subsection{Multiscale Blood Flow Solver}
\label{sec:multiscalesol}
We employ an experimentally validated blood flow solver~\cite{liu2018nanoparticle,liu2019nanoparticle,liu2019,liu2021computational} to verify our MSBM model for modelling cellular scale hallmarks in microcirculatory blood flows. The fluid phase is modeled using the three-dimensional lattice Boltzmann (LB) method~\cite{aidun2010lattice}. The RBC is modelled as a coarse-grained deformable capsule that adopts a biconcave discoidal shape under stress-free equilibrium, as outlined by Fedosov {\it et al.}~\cite{fedosov2010} and Pivkin and Karniadakis~\cite{pivkin2008}. The membrane nonlinear mechanics capture the RBC membrane viscoelasticity with physiological RBC shear elasticity, bending stiffness, and membrane viscosity~\cite{liu2021computational}. The RBC has a maximum diameter of 8 $\mu$m, and a membrane shear modulus (G) of $0.0063$ dyne cm$^{-1}$. The internal viscosity of RBC is set to five times that of plasma ($1.2$ cP). Periodic boundary condition in $x$ direction and no-slip boundary condition is applied at the wall. Pressure gradients are utilized to drive the flow.

\subsection{Quantification of Cell Free Layer}
In our analysis, we specify CFL as the near-wall region in which on average at most only half of one single RBC will fall inside the layer. Specifically, in all the simulations, we determine the CFL thickness by finding the near wall location, y=$\delta_{CFL}$, such that
\begin{equation}
     \phi(y=\delta_{CFL})=\frac{V_{RBC}}{2V(y)|y=\delta_{CFL}} \mbox{ ,} \label{eq:cfl}
\end{equation}
where $V_{RBC}$ is the volume of a single RBC with a value of $\sim$94 $\mu m^3$, $V(y)$ is the volume of the region spanning from the wall to $y$. As an example, for a tube with diameter 40 $\mu m$ as discussed in section~\ref{sec:tubularflow}, $\phi(y=\delta_{CFL})$ is calculated to be roughly $1\times10^{-4}$ as the threshold for us to determine the CFL thickness.

\section{Results} \label{sec:results}


\subsection{Parametric Study in Channel Flows}
\label{sec:2Dchannel}

\begin{figure}[h]
\begin{subfigure}{.5\textwidth}
  \centering
\includegraphics[width=.65\linewidth]{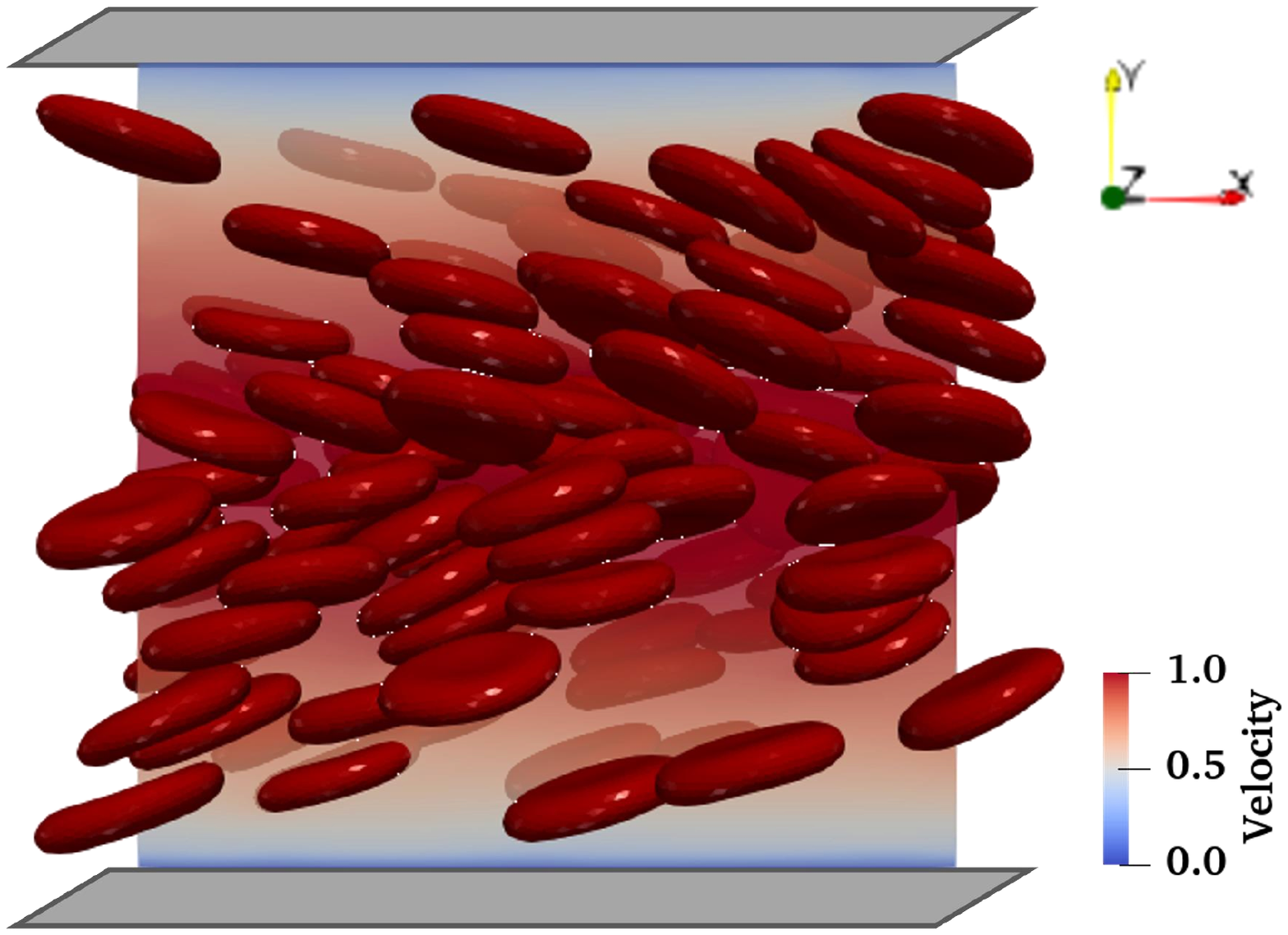}
\end{subfigure}%
\begin{subfigure}{.25\textwidth}
\centering

\tikzset{every picture/.style={line width=0.75pt}} 

\begin{tikzpicture}[x=0.72pt,y=0.75pt,yscale=-1,xscale=1]

\draw   (218.2,34.09) -- (439.8,34.09) -- (439.8,127.02) -- (218.2,127.02) -- cycle ;
\draw    (369.1,90.91) -- (369.13,62.19) ;
\draw [shift={(369.13,60.19)}, rotate = 90.06] [fill={rgb, 255:red, 0; green, 0; blue, 0 }  ][line width=0.08]  [draw opacity=0] (12,-3) -- (0,0) -- (12,3) -- cycle    ;
\draw    (369.1,90.91) -- (400.38,90.9) ;
\draw [shift={(402.38,90.9)}, rotate = 179.99] [fill={rgb, 255:red, 0; green, 0; blue, 0 }  ][line width=0.08]  [draw opacity=0] (12,-3) -- (0,0) -- (12,3) -- cycle    ;
\draw    (369.1,90.91) -- (387.02,74.31) ;
\draw [shift={(388.49,72.95)}, rotate = 137.21] [fill={rgb, 255:red, 0; green, 0; blue, 0 }  ][line width=0.08]  [draw opacity=0] (12,-3) -- (0,0) -- (12,3) -- cycle    ;
\draw  [color={rgb, 255:red, 0; green, 0; blue, 0 }  ,draw opacity=1 ][fill={rgb, 255:red, 0; green, 0; blue, 0 }  ,fill opacity=1 ] (218.2,119.59) -- (439.8,119.59) -- (439.8,127.02) -- (218.2,127.02) -- cycle ;
\draw    (447.71,121.3) -- (447.71,41.3) ;
\draw [shift={(447.71,39.3)}, rotate = 90] [fill={rgb, 255:red, 0; green, 0; blue, 0 }  ][line width=0.08]  [draw opacity=0] (12,-3) -- (0,0) -- (12,3) -- cycle    ;
\draw [shift={(447.71,123.3)}, rotate = 270] [fill={rgb, 255:red, 0; green, 0; blue, 0 }  ][line width=0.08]  [draw opacity=0] (12,-3) -- (0,0) -- (12,3) -- cycle    ;
\draw  [color={rgb, 255:red, 0; green, 0; blue, 0 }  ,draw opacity=1 ][fill={rgb, 255:red, 0; green, 0; blue, 0 }  ,fill opacity=1 ] (218.2,34.09) -- (439.8,34.09) -- (439.8,41.53) -- (218.2,41.53) -- cycle ;
\draw    (218.2,41.53) .. controls (243.86,45.13) and (289.75,87.73) .. (218.2,119.59) ;
\draw    (218.47,49.39) -- (234.43,49.39) ;
\draw [shift={(237.43,49.39)}, rotate = 180] [fill={rgb, 255:red, 0; green, 0; blue, 0 }  ][line width=0.08]  [draw opacity=0] (5.36,-2.57) -- (0,0) -- (5.36,2.57) -- cycle    ;
\draw    (218.47,58.62) -- (245.45,58.62) ;
\draw [shift={(248.45,58.62)}, rotate = 180] [fill={rgb, 255:red, 0; green, 0; blue, 0 }  ][line width=0.08]  [draw opacity=0] (5.36,-2.57) -- (0,0) -- (5.36,2.57) -- cycle    ;
\draw    (218.47,67.85) -- (251.87,67.85) ;
\draw [shift={(254.87,67.85)}, rotate = 180] [fill={rgb, 255:red, 0; green, 0; blue, 0 }  ][line width=0.08]  [draw opacity=0] (5.36,-2.57) -- (0,0) -- (5.36,2.57) -- cycle    ;
\draw    (218.47,77.79) -- (253.71,77.79) ;
\draw [shift={(256.71,77.79)}, rotate = 180] [fill={rgb, 255:red, 0; green, 0; blue, 0 }  ][line width=0.08]  [draw opacity=0] (5.36,-2.57) -- (0,0) -- (5.36,2.57) -- cycle    ;
\draw    (218.47,87.73) -- (252.79,87.73) ;
\draw [shift={(255.79,87.73)}, rotate = 180] [fill={rgb, 255:red, 0; green, 0; blue, 0 }  ][line width=0.08]  [draw opacity=0] (5.36,-2.57) -- (0,0) -- (5.36,2.57) -- cycle    ;
\draw    (218.47,96.96) -- (248.2,96.96) ;
\draw [shift={(251.2,96.96)}, rotate = 180] [fill={rgb, 255:red, 0; green, 0; blue, 0 }  ][line width=0.08]  [draw opacity=0] (5.36,-2.57) -- (0,0) -- (5.36,2.57) -- cycle    ;
\draw    (218.47,107.61) -- (236.27,107.61) ;
\draw [shift={(239.27,107.61)}, rotate = 180] [fill={rgb, 255:red, 0; green, 0; blue, 0 }  ][line width=0.08]  [draw opacity=0] (5.36,-2.57) -- (0,0) -- (5.36,2.57) -- cycle    ;
\draw    (221.2,148.47) -- (439.63,148.47) ;
\draw [shift={(441.63,148.47)}, rotate = 180] [fill={rgb, 255:red, 0; green, 0; blue, 0 }  ][line width=0.08]  [draw opacity=0] (12,-3) -- (0,0) -- (12,3) -- cycle    ;
\draw [shift={(218.2,148.47)}, rotate = 0] [fill={rgb, 255:red, 0; green, 0; blue, 0 }  ][line width=0.08]  [draw opacity=0] (8.93,-4.29) -- (0,0) -- (8.93,4.29) -- cycle    ;

\draw (399.75,84.89) node [anchor=north west][inner sep=0.75pt]  [font=\tiny]  {$\mathit{x}$};
\draw (361.55,50.41) node [anchor=north west][inner sep=0.75pt]  [font=\tiny]  {$y$};
\draw (383.86,62.96) node [anchor=north west][inner sep=0.75pt]  [font=\tiny]  {$z$};
\draw (447.99,68.44) node [anchor=north west][inner sep=0.75pt]    {$\mathit{H}$};
\draw (217.37,128.54) node [anchor=north west][inner sep=0.75pt]  [font=\scriptsize]  {$u_{x}( y=0) \ =0$};
\draw (212.07,16.77) node [anchor=north west][inner sep=0.75pt]  [font=\scriptsize]  {$u_{x}( y=H) \ =0$};
\draw (322.28,148.84) node [anchor=north west][inner sep=0.75pt]    {$L$};
\draw (257.8,67.54) node [anchor=north west][inner sep=0.75pt]  [font=\small]  {$u_{x}( y)$};
\draw (188.61,119.59) node [anchor=north west][inner sep=0.75pt]  [font=\scriptsize,rotate=-268.79]  {$\phi ( x=0) =\phi _{b}$};
\draw (200.33,119) node [anchor=north west][inner sep=0.75pt]  [font=\scriptsize,rotate=-270.4]  {$u_{y}( x=0) =0$};
\draw (297.72,16.61) node [anchor=north west][inner sep=0.75pt]  [font=\scriptsize]  {$\partial \phi /\partial y=0$};
\draw (303.33,130.11) node [anchor=north west][inner sep=0.75pt]  [font=\scriptsize]  {$\partial \phi /\partial y=0$};
\draw (453.22,42.44) node [anchor=north west][inner sep=0.75pt]  [font=\scriptsize]  {$\partial \phi /\partial x=0$};
\draw (451.22,93.1) node [anchor=north west][inner sep=0.75pt]  [font=\scriptsize]  {$\partial u_{x} /\partial x=0$};
\draw (365.55,14.83) node [anchor=north west][inner sep=0.75pt]  [font=\scriptsize]  {$\mathbf{j}_{\perp }^{\prime } \cdotp \mathbf{n} =0$};
\draw (475.48,68.14) node [anchor=north west][inner sep=0.75pt]  [font=\scriptsize]  {$p=0$};
\draw (375.58,127.33) node [anchor=north west][inner sep=0.75pt]  [font=\scriptsize]  {$\mathbf{j}_{\perp }^{\prime } \cdotp \mathbf{n} =0$};

\end{tikzpicture}
\end{subfigure}%
\caption{\label{fig:channelgeom} Blood flows through a microchannel. The left snapshot shows the blood flow through a channel obtained using our in-house multiscale blood flow solver. The schematic on the right shows the numerical setup for the MSBM simulation of blood flow through a channel. We consider a channel width of H=$42.7 \, \mu$m height channel with bulk hematocrit $\phi_b=0.2$.}
\end{figure}

Here we describe the planar channel-flow geometry that is used to simulate the pressure-driven flow of blood in a channel of rectangular cross-section (see figure~\ref{fig:channelgeom}). Using a Cartesian coordinate system, where $x$ is the flow direction, $y$ is the gradient direction and $z$ is the neutral direction, the channel dimensions are $H \times L \times W$, where $H$ is the height, $L$ is the length, and $W$ is the width. 

To initialize the simulation, at $t=0$, the field velocity is zero ($\bun =  \bZ$), the stress tensors are the zero tensor (i.e. $\btaun = \bZ$), the pressure is zero and the volume fraction in the whole channel is equal to the bulk volume fraction (or the hematocrit) $\phi_b$. 

At the inlet, we set a velocity profile $u_x(y)$ that will adopt a parabolic shape of the form $u_x(y)=U_{0}\,(H-y)\,y/(H/2)^2$; for the volume fraction, we impose the value of the hematocrit ($\phi_b$), and zero gradient boundary conditions for both the pressure and the flux vector $\bm{j}_{\perp}^{\, \prime}$. At the walls ($y=0$ and $y=H$), we have no-slip boundary conditions for the velocity field, zero gradient boundary conditions for both the pressure and the volume fraction fields and for the vector $\bm{j}_{\perp}^{\, \prime}$ (which has units of velocity) we impose a $\textit{slip}$ boundary condition, which physically enforces zero mass flux through the wall by setting zero value to the normal velocity component. Finally, at the outlet, we set fully developed boundary conditions (i.e. zero value for pressure and zero-gradient boundary conditions for the velocity field and for the volume fraction).

The time step $\Delta t$ used in the simulations depends on the mesh and on the selected model parameters, but we always ensure that the Courant number condition ($Co=\Delta t \, U_{0}/\Delta x \ll 1$, where $\Delta x$ is the value of the cell size) is always satisfied.

To satisfy the zero-inertia conditions seen in blood flows, a particle Reynolds number $Re_p$ and a global Reynolds number for channel flow were defined following Dbouk~{\it et al.}~\cite{dbouk}, which are $Re_p=4/3 \,\rho \,a^3\,U_{0}/\big[\eta \,(H/2)^2 \big]$ and $Re=\rho \, U_{0} \,H/\eta$, where $\eta=\mu_f\,\mu_r(\phi)$. As these Reynolds numbers are typically much smaller than one for microcirculatory flows, we reasonably neglect the inertia. 

The aspect ratio $L/H$ of the channel has to be carefully chosen to ensure that the flow profiles from our simulations are fully developed. Nott and Brady~\cite{Nott1994} derived an equation that guarantees this condition provided that:
\begin{equation}
\label{eq:aspectratio}
\Big(\frac{L}{H}\Big) \geq \frac{1}{12 g(\phi)} \Big(\frac{H}{a}\Big)^2 \qquad g(\phi) = \frac{1}{3} \Big( 1+\frac{1}{2} \exp{(8.8 \phi)}  \Big).
\end{equation}

The velocity and other flow quantities will depend on the coordinate $y \in [0,H]$ ($H$ being the channel-height) and since we will be integrating the governing equations from $y=0$ to $y=H$ and along the $x$-direction, we consider the effect of the two-walls on the blood flow. Thus, the lift term becomes:

\begin{equation}
    \bm{L}_{\perp} = \frac{3\,\mu_f \,\gd}{4 \,\pi \, (y+h_0)^{\beta}} \, f(1-\nu) \, (0,1,0) +  \frac{3\,\mu_f \,\gd}{4 \,\pi \, (H-y+h_0)^{\beta}} \, f(1-\nu) \, (0,-1,0)\mbox{ .} \label{eq:lift_force12}
\end{equation}

\subsubsection{Comparison with classical SBM}
\label{sec:msbmvssbm}

\begin{figure}[hbt!]
\begin{subfigure}{.5\textwidth}
  \centering
  \includegraphics[width=.9\linewidth]{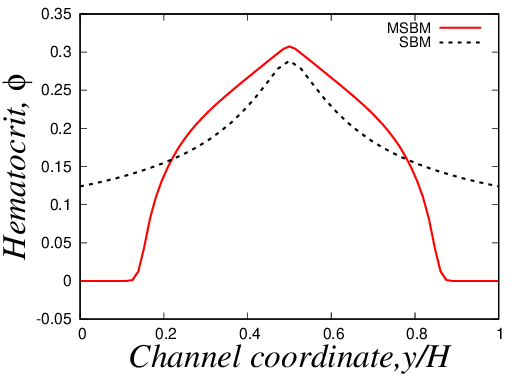}
  \caption{$\phi$ vs $y/H$}
  \label{fig:hemavsySBMMSBM}
\end{subfigure}%
\begin{subfigure}{.5\textwidth}
  \centering
  \includegraphics[width=.9\linewidth]{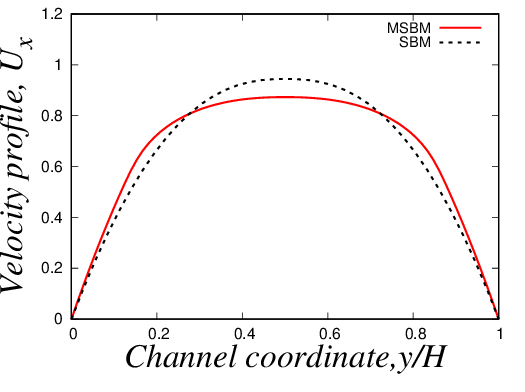}
  \caption{$u_x$ vs $y/H$}
  \label{fig:UxvsySBMMSBM}
\end{subfigure}%
\caption{\label{fig:BMSBMvsMSBM} Comparison between the original suspension balance model (SBM) and our modified suspension balance model (MSBM). For both simulations, we use a channel height of $H=50\,\mu$m, the following parameters are used: $\mu_f/\rho= 1.30 \times 10^{-6}$ m$^2$/s, $\phi_m=0.5$, $\phi_b=0.2$, $a=2.82 \, \mu$m. For our MSBM simulations, we use $\alpha=4$, $f(1-\nu)=1.2$, $\beta=1.2$ and $h_0=1.0\times 10^{-12}$ m.  On the left, we show the hematocrit vs channel coordinate profiles; on the right, we illustrate the velocity vs channel coordinate profiles. The solid red lines show the MSBM results, while the black dashed lines correspond to the SBM predictions.}
\end{figure}

In this section, we illustrate the differences between the classical suspension balance model (SBM) and our recently developed suspension balance model (MSBM) for blood flows, where we carry out channel flow simulations of particles with radius $a=2.82 \, \mu$m corresponding to the hydrodynamic radius of an RBC, a bulk volume fraction concentration of $\phi_b=0.2$, which is a commonly encountered value in blood flow in microvessels~\cite{liu2019}, and a maximum packing concentration of $\phi_m=0.5$. $U_0=0.01$ m/s is imposed at the inlet, which is ideal to obtain a physiologically relevant shear rate of $\gd \approx 1000$ s$^{-1}$ at the walls. The plasma viscosity is $\mu_f=0.0013$ Pa s and the density is $\rho=1000$ kg/m$^3$. 
Our results can be found in figure~\ref{fig:BMSBMvsMSBM}, where it is evident that both models predict particle migration from regions of high stress to low stress, where blood cells move away from the walls toward the center of the geometry ($y=H/2$). This leads to high volume fraction in the region where the shear rate $\gd$ is negligible near the centerline. On the other hand, the MSBM predicts the CFL phenomenon, where a zero concentration of RBC is observed near the channel walls (at $y=0$ and at $y=1$), while a nonzero volume fraction at the channel-walls is observed for the conventional SBM. This leads us to conclude that the latter model fails to capture the CFL near the walls, and thus, the addition of the lift force was needed for the continuum model to capture this phenomenon. For the velocity profiles (see figure~\ref{fig:UxvsySBMMSBM}), we can notice that the SBM predicts a parabolic-like profile for the suspension, while our MSBM predicts a plug-flow like for blood~\cite{roman2012}. These kinds of flow profiles commonly appear in two-phase flows, shear-banding flows and yield-stress fluids~\cite{castisanc2022}, where there exists a rich region of a high-viscosity (or a solid-like behavior for yield-stress materials) at the centerline, while the regions near the wall are dominated by a low-viscosity (or a viscous region for yield-stress materials). The high volume fraction is responsible for blunting or flattening the parabolic velocity profile at the core of the channel. Our results agree with observations made by Lyon and Leal~\cite{Lyon1998}.



\subsubsection{Temporal effect}
\label{sec:temp}

In this section, we study the temporal evolution of both the hematocrit and velocity profiles predicted by our recently developed modified suspension balance model for blood flows. 
 Here we simulate RBCs flowing in a channel of $42.7 \, \mu$m height, and set the parameters of our modified suspension balance model 
 is: $\mu_f/\rho= 1.30 \times 10^{-6}$ m$^2$/s, $\phi_m=0.50$, $\phi_b=0.25$, $\beta=1.2$, $\alpha=4$, $a=2.82 \, \mu$m, $f(1-\nu)=1.2$, $h_0=1.0\times 10^{-12}$ m and a value of $\epsilon = a/H=0.13$. Since the global and particle Reynolds numbers for these simulations are $Re=0.041$ and $Re_p=1.26 \times 10^{-4}$, respectively, the temporal and convective terms of the momentum conservation can be ignored. We also carried out a full-transient simulation (not shown here) and the results obtained are identical to the ones reported here.


\begin{figure}[hbt!]
\begin{subfigure}{.5\textwidth}
  \centering
  \includegraphics[width=.9\linewidth]{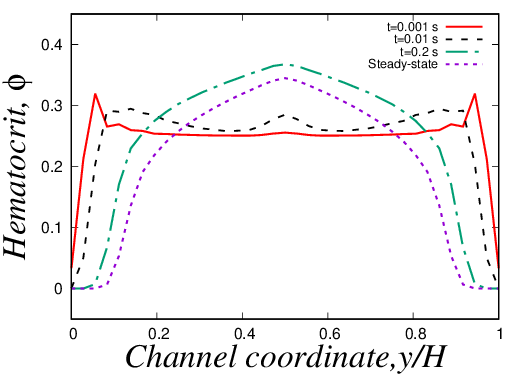}
  \caption{$\phi$ vs $y/H$}
  \label{fig:hemavstvar}
\end{subfigure}%
\begin{subfigure}{.5\textwidth}
  \centering
  \includegraphics[width=.9\linewidth]{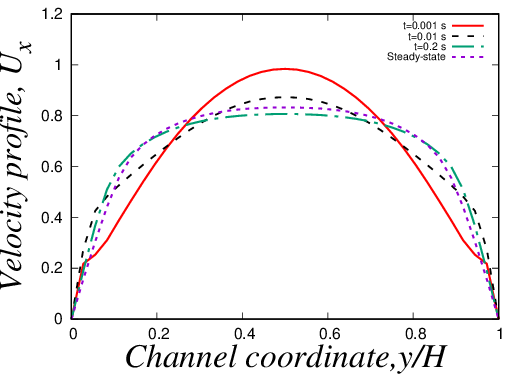}
  \caption{$u_x$ vs $y/H$}
  \label{fig:Uxvstvar}
\end{subfigure}%
\caption{\label{fig:tvarprofiles} Temporal evolution of hematocrit (left) and velocity (right) profiles in a channel of $42.7 \, \mu$m height. The red solid line represents the snapshot for $t=1 \times 10^{-3}$ s, the black dashed line is for the case $t=1 \times 10^{-2}$ s, the green dashed-dotted line represents the case for $t=0.2$ s and the purple dashed line is the steady-state solution. The global and particle Reynolds numbers for these simulations are $Re=0.041$ and $Re_p= 1.26 \times 10^{-4}$, respectively.}
\end{figure}

Our transient results can be found in figure~\ref{fig:tvarprofiles}, where a uniform hematocrit profile is applied to match the bulk hematocrit, $\phi_b$. 
Four different snapshots are shown: at the very beginning of the simulation ($t=1 \times 10^{-3}$ s), which is represented by solid red lines, we can see that the volume fraction profile (figure~\ref{fig:hemavstvar}) remains flat near the center, and the concentration peak is close to the wall~\cite{qi2018time}. Although $\phi$ decreases as we approach the walls, our model does not immediately predict a cell-depleted layer, in contrast to the transient predictions of Qi and Shaqfeh~\cite{qi2018time}. For the velocity profile at this time (see figure~\ref{fig:Uxvstvar}), we observe a Newtonian-like parabolic profile due to the mostly uniform hematocrit distribution. Around $t=1 \times 10^{-2}$ s (black dashed lines) we see the formation of a peak of $\phi$ in the center, while the peaks that were initially observed at the beginning of the simulation become less sharp and decrease at this time. On the other hand, the velocity starts to lose its parabolic shape and it widens near the wall. 
It is around $t \approx 0.2$ s (green dashed-dotted line) when we start to observe the formation of the CFL near the walls and the peak concentration at the centerline, meanwhile for the velocities, we notice that the profile is no longer parabolic and starts to take the plug-flow like form discussed in previous sections. Finally, the flow evolves until the steady-state solution is reached, which is represented by the purple dashed lines.


It is also important to point out that the average hematocrit $\hat{\phi}$ (computed as $\hat{\phi} =\int_{0}^{H} \big[\phi (y) \big]  \,dy/H$) is smaller compared to the bulk hematocrit value $\phi_b$ imposed at the inlet, and this is more evident when comparing the hematocrit profiles when $t=0.4$ s and when steady-state is reached, where there is a decrease of the average hematocrit $\hat{\phi}$. Miller {\it et al.}~\cite{Miller2006} explained this as the need to maintain a constant flux of particles at any axial station. However, we note that this phenomenon is in fact well known as the Fåhræus Effect in the blood rheology community~\cite{pries1990}. The Fåhræus Effect describes the discharged hematocrit (defined as the bulk hematocrit here) is greater than the tube hematocrit (calculated as average hematocrit here) as a result of the faster-moving red blood cells in the center of the vessel being discharged more compared to the slower-moving plasma at the circumference (more discussed in \S~\ref{sec:fahraeuseff}). In fact, from mass conservation for the RBC phase, a velocity weighted “average” hematocrit (or equivalently the discharge hematocrit)  $\hat{\phi_w}$ can be computed as $\hat{\phi_w} =\int_{0}^{H} \big[\phi \, U_x(y) \big]  \,dy/\int_{0}^{H} U_x(y)  \,dy$. It has been confirmed in our simulation that $\hat{\phi_w}\approx\phi_b$ always holds with a less than 3\% numerical error at any axial location and at any time point of the simulation.

\subsubsection{Lift force effect}
\label{sec:betavar}

Different deformable particles exhibit different capacities in breaking the hydrodynamic symmetry, which causes variation in the lift force dependency on the particle-wall distance. For instance, RBCs with a biconcave disc shape and higher deformability~\cite{reasor2013determination} typically exhibit a higher lift force near the wall compared to typical vesicles~\cite{Abkarian2002}. To capture this lift force variability and also identify a suitable parameter of blood flows, we study the effect of the $\beta$ parameter on the changes of hematocrit distribution and CFL thickness.
For our simulations, we use a uniform Cartesian mesh with the following dimensions: $L=0.004$ m, $H=50 \,\mu$m, $W=10 \, \mu$m, using $100 \times 60 \times 1 $ control volumes. 
We set particle size equal to the RBC hydrodynamic radius $a=2.82 \, \mu$m. The value of $U_0$ imposed at the inlet is $U_0=0.01$ m/s. The plasma viscosity is $\mu_f=0.0013$ Pas, the density is $\rho=1000$ kg/m$^3$, the bulk hematocrit is $\phi_b=0.2$. 

\begin{figure}[hbt!]
\begin{subfigure}{.5\textwidth}
  \centering
  \includegraphics[width=.9\linewidth]{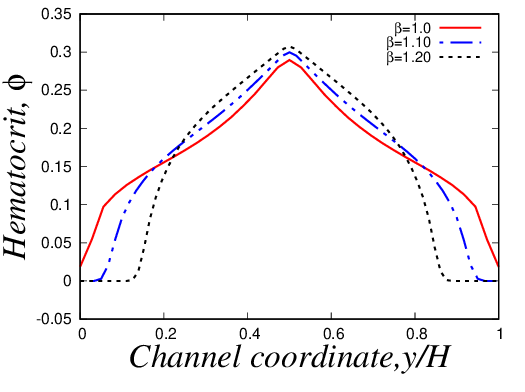}
  \caption{$\phi$ vs $y/H$}
  \label{fig:hemavsbetavar}
\end{subfigure}%
\begin{subfigure}{.5\textwidth}
  \centering
  \includegraphics[width=.9\linewidth]{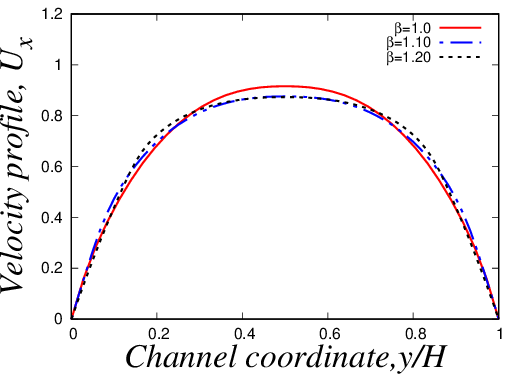}
  \caption{$u_x$ vs $y/H$}
  \label{fig:Uxvsbetavar}
\end{subfigure}%
\caption{\label{fig:betavarprofiles} Hematocrit (left) and velocity (right) profiles obtained using our $\textit{SbmBloodFoam}$ solver for different values of the coefficient $\beta$. The blood parameters used in our simulations are: $\mu_f/\rho= 1.30 \times 10^{-6}$ m$^2$/s, $\phi_m=0.5$, $\phi_b=0.2$, $\alpha=4$, $a=2.82 \, \mu$m, $f(1-\nu)=1.2$ and $h_0=1.0\times 10^{-12}$ m. The solid red line corresponds to a value of $\beta=1.0$, the blue solid-dashed line is for $\beta=1.10$ and the black dashed line is for the case with $\beta=1.20$.}
\end{figure}

Our results can be found in figure~\ref{fig:betavarprofiles}, where we illustrate our steady-state hematocrit and velocity profiles obtained at $x/L=0.875$ for three different values of the coefficient $\beta=[1.0,1.1,1.2]$. We limit ourselves to these values, since we have found that a $\beta$ value within this range is able to reproduce behavior observed in blood flows. As can be seen in figure~\ref{fig:hemavsbetavar}, we illustrate the behavior of the hematocrit profiles as we vary the $\beta$ parameter. Since an increase in value of $\beta$ leads to a higher magnitude of the lift force, it can be easily noticed that the CFL thickness $\delta_{CFL}$ rises as $\beta$ is being increased. As it can be noticed, no significant CFL is observed when $\beta=1.0$, since the hematocrit is non-zero at the wall, but when $\beta=1.10$ we observe a CFL with a thickness value of $\delta_{CFL}\approx 2.4\,\mu$m, and when $\beta=1.20$ we obtain a value of $\delta_{CFL}\approx 6.2\,\mu$m. This leads us to conclude that $\beta$ controls the CFL thickness in the simulations. It is also worth noting that as $\beta$ increases, we observe higher values of hematocrit at the center of the channel as more RBCs migrate from the wall region to the cell-laden region. 

We also report the corresponding velocity profiles for these simulations, which can be found in figure~\ref{fig:Uxvsbetavar}), where a parabolic-like profile is obtained with a semi-flat region at the center. As we approach the walls, which is where the cell-depleted region is found, the velocity values decrease until they reach a zero value, as specified by our no-slip boundary conditions. The only noticeable differences between the three $\beta$ cases are observed outside the centerline, where slightly higher values of velocity are seen when the value of $\beta$ has the greatest value. Overall, the velocity is fairly insensitive to $\beta$.

\subsubsection{Bulk hematocrit effect}
\label{sec:hemavar}

In this section, we study the effect of varying the bulk hematocrit $\phi_b$. The bulk hematocrit considered here range from 0.1$\sim$0.32, as we focus on blood flow in microcirculation where the average bulk hematocrit is generally lower than the systemic hematocrit (0.4$\sim$0.45)~\cite{barbee1971fahraeus,klitzman1979microvascular,gaehtgens1981}. In microvescular networks when Zweifach-Fung effects~\cite{liu2020heterogeneous} are significant, transient superphysiological hematocrit could be observed \cite{pries1990} and is not considered here. However, technically, when simulating high bulk hematocrit conditions ($>$0.3), as the CFL thickness gets smaller, spatiotemporal resolution needs to be increased significantly to capture the sharp transition from the CFL to the RBC-laden region.



\begin{figure}[hbt!]
\begin{subfigure}{.5\textwidth}
  \centering
  \includegraphics[width=.9\linewidth]{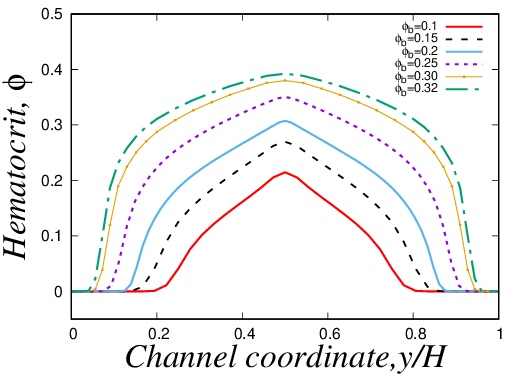}
  \caption{$\phi$ vs $y/H$}
  \label{fig:hemavsphibvar}
\end{subfigure}%
\begin{subfigure}{.5\textwidth}
  \centering
  \includegraphics[width=.9\linewidth]{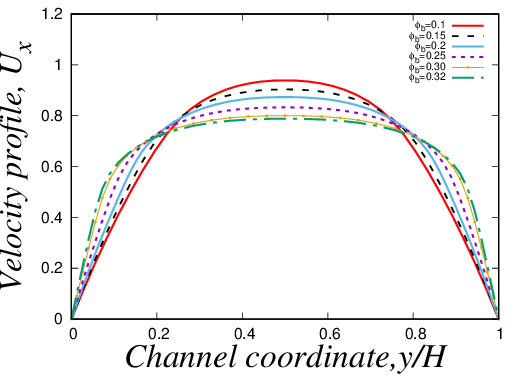}
  \caption{$u_x$ vs $y/H$}
  \label{fig:Uxvsphibvar}
\end{subfigure}%
\caption{\label{fig:phibvarprofiles} Hematocrit (left) and velocity (right) profiles obtained using our $\textit{SbmLiftFoam}$ solver for different values of bulk hematocrit $\phi_b$. These curves were obtained in channels of $50 \, \mu$m height. The blood parameters used in our simulations are: $\mu_f/\rho= 1.30 \times 10^{-6}$ m$^2$/s, $\phi_m=0.5$, $\beta=1.20$, $\alpha=4$, $a=2.82 \, \mu$m, $f(1-\nu)=1.2$ and $h_0=1.0\times 10^{-12}$ m. From top to bottom: $\phi_b=0.32, 0.30, 0.25, 0.20, 0.15, 0.1$.}
\end{figure}

In figure~\ref{fig:hemavsphibvar}, we show our hematocrit profiles, where the curves at the top have highest values of $\phi_b$. For all hematocrits, we observe similar tendencies: the formation of a two-phase flow, where a core with higher concentrations of RBC is formed at the centerline and a cell-depleted fluid zone near the walls. In addition, a more pronounced peak is seen for the profiles with low bulk hematocrit values (see $\phi_b=0.1, 0.15$ and $0.2$ cases), while a more rich zone of RBCs near the channel-center is seen for the cases with high hematocrit (see $\phi_b=0.3$ and $0.32$ cases). Interestingly, it can be noticed that the CFL thickness tends to decrease as the value of bulk hematocrit increases, which is consistent with results reported in the literature (see figure~\ref{fig:CFLvsHemaExpSimA} and~\cite{vahidkhah2014,salame2025}).

The velocity profiles for these cases are illustrated in figure~\ref{fig:Uxvsphibvar}, where we observe a plug flow zone at the center for the cases with high values of $\phi_b$, which is in agreement with observations made by Karnis {\it et al.}~\cite{karnis1966}. As the value of the bulk hematocrit decreases, this plug-flow core lessens, leading to a semi-parabolic velocity profile.

\subsubsection{Maximum packing concentration effect}
\label{sec:maxpack}
\begin{figure}[hbt!]
\begin{subfigure}{.5\textwidth}
  \centering
  \includegraphics[width=.9\linewidth]{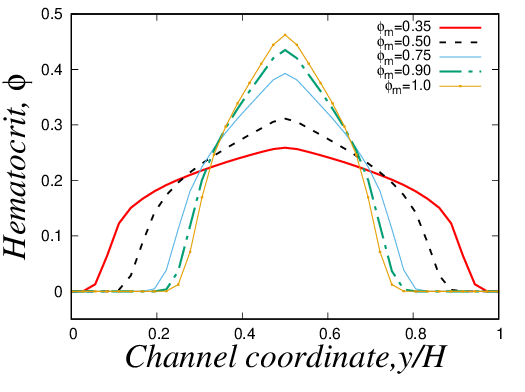}
  \caption{$\phi$ vs $y/H$}
  \label{fig:hemavsphimvar}
\end{subfigure}%
\begin{subfigure}{.5\textwidth}
  \centering
  \includegraphics[width=.9\linewidth]{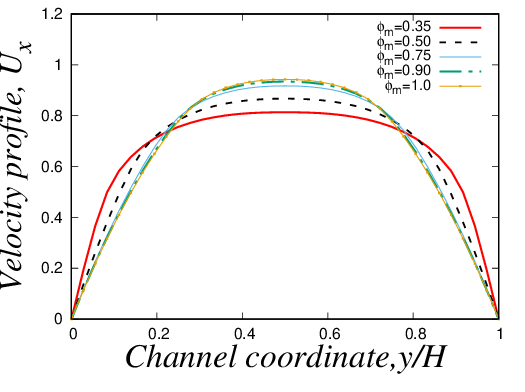}
  \caption{$u_x$ vs $y/H$}
  \label{fig:Uxvsphimvar}
\end{subfigure}%
\caption{\label{fig:phimvarprofiles} Hematocrit (left) and velocity (right) profiles obtained using our $\textit{SbmLiftFoam}$ solver for different values of maximum packing concentration $\phi_m$. These curves were obtained in channels of $50 \, \mu$m height. The blood parameters used in our simulations are: $\mu_f/\rho= 1.30 \times 10^{-6}$ m$^2$/s, $\phi_b=0.2$, $\beta=1.20$, $\alpha=4$, $a=2.82 \, \mu$m, $f(1-\nu)=1.2$ and $h_0=1.0\times 10^{-12}$ m. Red solid line is for the case $\phi_m=0.35$, black dashed line for $\phi_m=0.50$, blue thin line for $\phi_m=0.75$, dotted-dashed line for $\phi_m=0.90$ and solid-dotted line for $\phi_m=1.0$.}
\end{figure}

RBC local packing (or local hematocrit) depends on not only the bulk hematocrit but also the RBC-RBC hydrodynamic interactions (lubrication forces) and RBC-RBC adhesion/aggregation (e.g., sickle cell disease or diabetes blood enhances aggregation~\cite{liu2021computational}). In the MSBM we are developing, the maximum packing concentration $\phi_m$ can be thought of as a semi-empirical parameter that adjusts the RBC local packing propensity in the flow as a result of RBC-RBC. Using channels of $50 \, \mu$m height, we report our results in figure~\ref{fig:phimvarprofiles}, where five cases were simulated using different values of $\phi_m:0.35, 0.50,0.75, 0.9$ and $1.0$, where the global $Re$ and particle $Re_p$ Reynolds numbers values for these simulations lie in the following ranges: $Re=0.035-0.12$ and $Re_p=6.76 \times 10^{-5}-2.36 \times 10^{-4}$. 

From the hematocrit profiles (see figure~\ref{fig:hemavsphimvar}), it is evident that the maximum packing concentration has a strong influence on the shear-induced particle migration and on the CFL thickness. For instance, for the case with the lowest value of $\phi_m=0.35$, we observe the formation of a thin CFL near the wall ($\delta_{CFL}\approx 1.4 \, \mu$m) and a maximum hematocrit value equal to $0.26$. However, as we increase the maximum packing concentration, higher hematocrits can be achieved near the centerline since the RBCs can still be packed due to the freedom provided by the high values of $\phi_m$, leading to RBC aggregation. Under these conditions that favor particle migration, we also observe thicker cell-depleted regions; for instance, for the case with $\phi_m=1.0$, we observe a maximum hematocrit value of $0.46$ and a CFL thickness value of $\delta_{CFL}\approx 11.1 \, \mu$m. The velocity profiles are shown in figure~\ref{fig:Uxvsphimvar}, and it can be seen that a decrease in $\phi_m$ flattens the velocity profile near the centerline. This is because the peak hematocrit is less pronounced in this region compared to the high $\phi_m$ cases, where the velocity profile adopts a semi-parabolic shape. Notably, the higher CFL thickness predicted with higher $\phi_m$ in our simulation is consistent with the literature showing higher RBC aggregation leads to thicker CFL~\cite{ong2010effect}, confirming $\phi_m$ does capture the RBC packing/aggregation at least qualitatively and can be thought of as a semi-empirical parameter that controls RBC local packing propensity. Overall, we found $\phi_m$=0.5 appears to best reflect the normal hemorheology as we show in the following sections.


\subsection{Comparing to Existing Theory and Experiments}
\label{sec:theoryvsexp}

\subsubsection{Lift-induced RBC Migration under Simple Shear}
\label{sec:scaling}

Grandchamp {\it et. al.}~\cite{grandchamp2013} studied the scaling behaviors of RBC lift-induced migration away from the wall. They quantified the RBC-wall distance $z$ as a function of $t\,\gd$, where they observed a linear scaling relation, 
\begin{equation}
    	z^3 \sim 3 \, U\,R^3\,\gd \, t \label{eq:scaling}
\end{equation}
where $R$ is the particle characteristic size and $U$ is a dimensionless drift velocity that depends on the particle shape. Since the cell-free layer is essentially formed by the same lift effect that causes collective RBC migration away from the wall, we hypothesize that the CFL thickness, $\delta_{CFL}$, should follow similar temporal scaling behavior in short-time scales in semi-dilute concentrations when before cell-cell interactions become significant.

\begin{figure}[hbt!]
\begin{subfigure}{.5\textwidth}
  \centering
  \includegraphics[width=.9\linewidth]{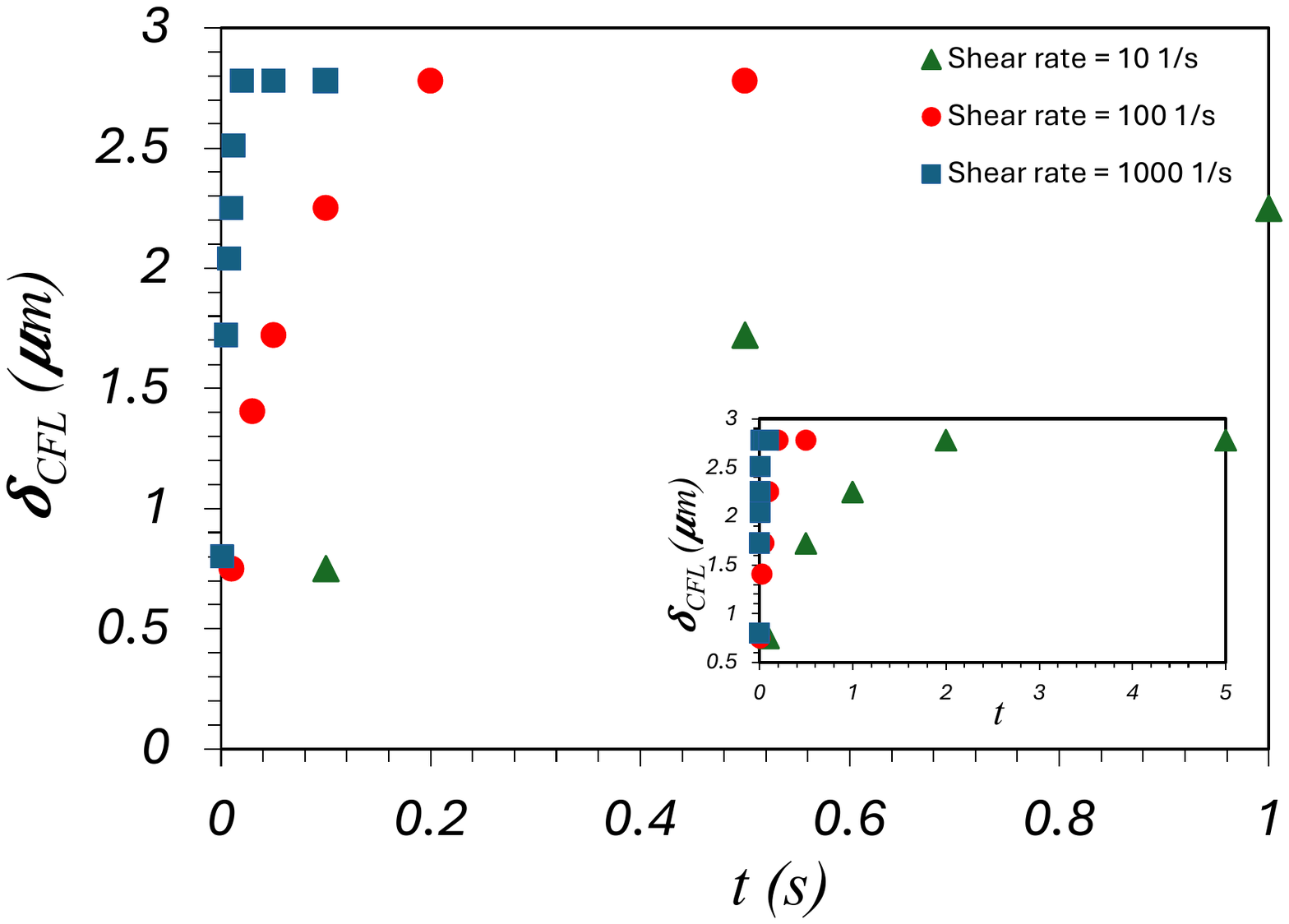}
  \caption{$\delta_{CFL}$ vs $t$}
  \label{fig:ssfunscaled}
\end{subfigure}%
\begin{subfigure}{.5\textwidth}
  \centering
  \includegraphics[width=.9\linewidth]{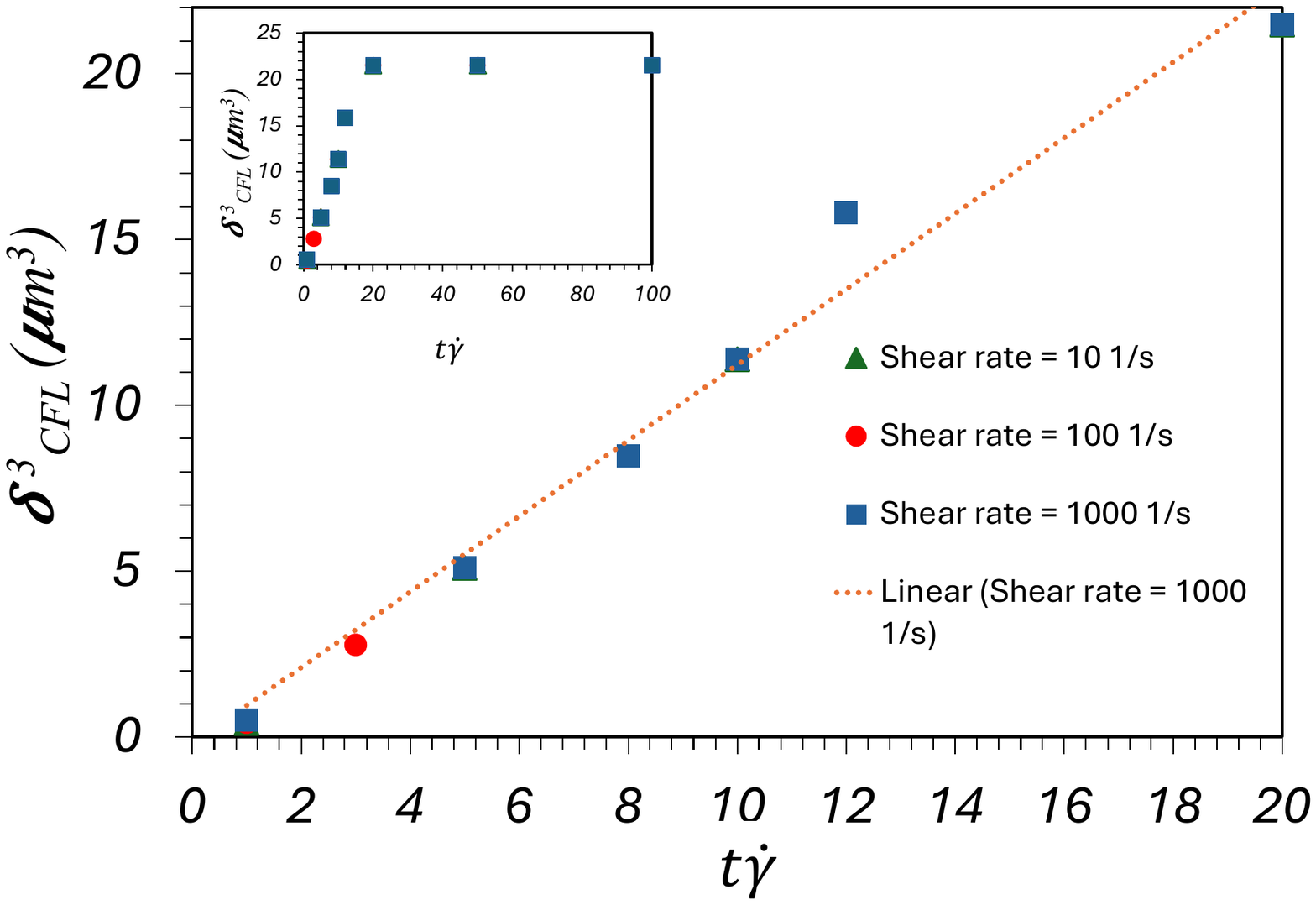}
  \caption{$\delta_{CFL}^3$ vs $t\,\gd$}
  \label{fig:ssf}
\end{subfigure}%
\caption{\label{fig:shearflowdelta3tgd} Wall-RBC distance quantified by the cell free layer as a function of time in simple shear flow with $\phi_b=0.15$ fixed. Figure~\ref{fig:ssfunscaled}) shows the unscaled behavior of $\delta$ vs $t$ (inlet shows results for longer times), while figure~\ref{fig:ssf}) illustrates the curve for the cubic of CFL thickness $\delta_{CFL}^3$ vs dimensionless time $t \,\gd$, where the inlet figure shows the linear behavior at small times. The green triangles are for the case with $\gd=10$ s$^{-1}$, red circles correspond to the case with $\gd=100$ s$^{-1}$, and blue squares were obtained with $\gd=1000$ s$^{-1}$.}
\end{figure}

To test this hypothesis, we conducted simulations of wall-bounded simple shear flow of blood flows using the MSBM. 
We simulate a channel with the following dimensions: height $H=2.7 \times 10^{-5}$ m, length $L=0.004$ m and width $W=1\times 10^{-5}$ m. The bottom wall is fixed (with non-slip boundary condition for the velocity) and the top wall is moving with a constant velocity $U$ in $x$-direction. Zero-gradient boundary conditions for the pressure, velocity and the volume fraction are imposed at the east, west and bottom walls. We fix the value of bulk hematocrit $\phi_b=0.15$ and we simulate simple shear flows using three different values of shear rate $\gd = 10, 100$ and $1000$ s$^{-1}$. The MSBM parameter values used are: $\mu_f/\rho= 1.32 \times 10^{-6}$ m$^2$/s, $\beta=1.2$, $\phi_m=0.5$, $\alpha=4$, $a=2.82 \, \mu$m, $f(1-\nu)=1.2$ and $h_0=1.0\times 10^{-12}$ m.

Figure~\ref{fig:ssfunscaled} illustrates the behavior of the CFL thickness as a function of time, where we see that the CFL is formed much quicker for flows with higher shear rate. As CFL gets thicker and cell-cell interations in the cell-laden region becomes significant, the migration rate (slope) decreases and eventually plateaus. In figure~\ref{fig:ssf}, we rescale the data and plot the cubic of CFL thickness $\delta_{CFL}^3$ vs dimensionless time $t \,\gd$, where we see a good collapse of $\delta_{CFL}^3$-$t \,\gd$ under various shear rates (see the inset of Fig~\ref{fig:ssf}. More interestingly, at small time scales ranging from $t \,\gd$=0 to 20, a linear scale, $\delta_{CFL}^3 \sim \, \gd \,t $, is observed similar to the single RBC lift-induced migration scaling reported experimentally by Grandchamp et al.~\cite{grandchamp2013}. The overall good collapse and linear scaling in the CFL evolution confirm the MSBM equipped with the lift forcing term seems to capture the essence of the lift-induced RBC migration effect at single cellular level.

\subsubsection{Compared to Experimental Blood Flows}
\label{sec:exp}

\begin{figure}[hbt!]
  \centering
  \includegraphics[width=.7\linewidth]{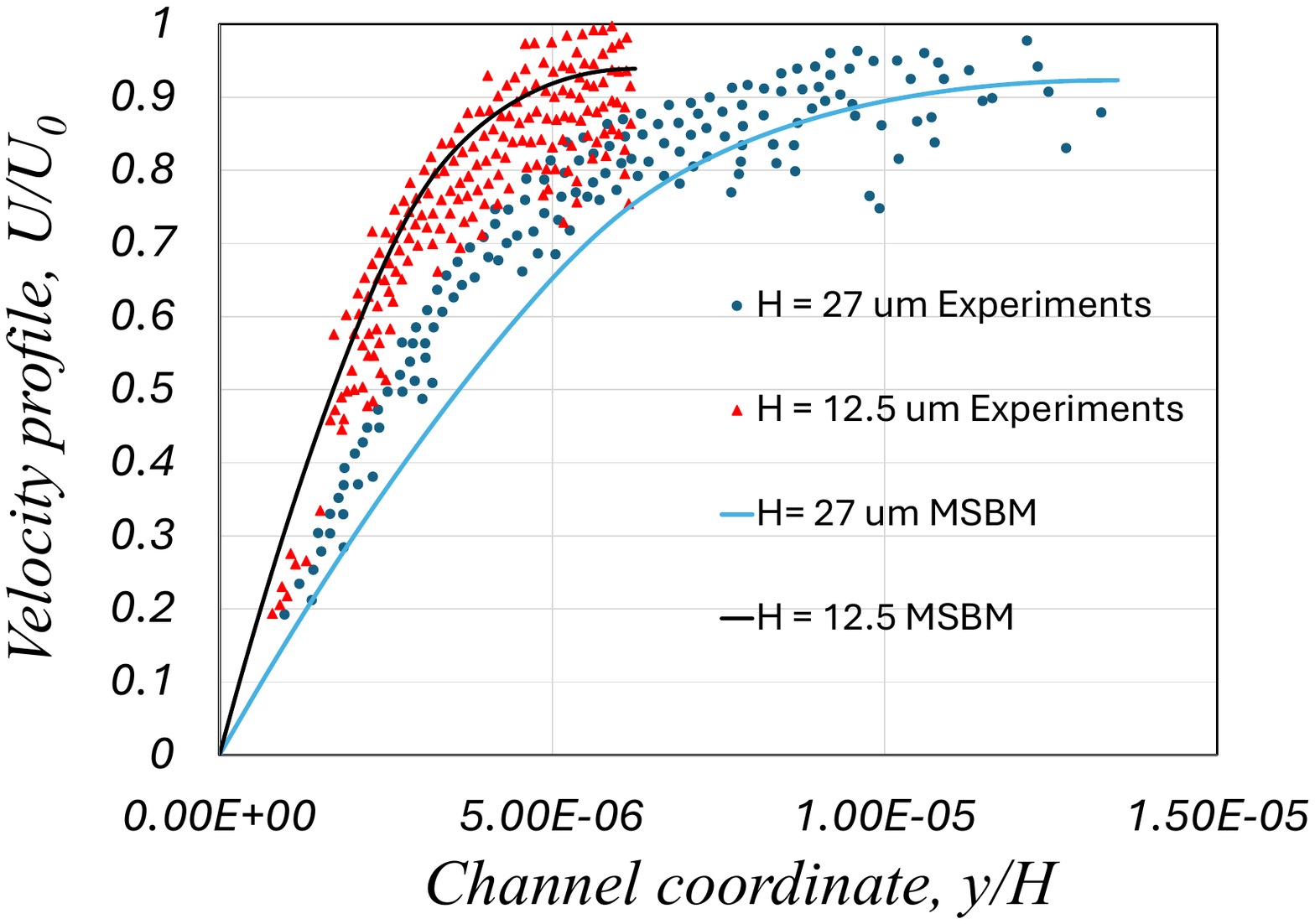}
\caption{\label{fig:losserand} Comparison between MSBM predictions (solid lines) and experimental results (dots) reported by Losserand {\it et. al.}~\cite{losserand2019}, who measured experimentally the velocity profile of RBCs in different channel sizes with a fixed bulk concentration of 0.1. The MSBM parameter values used for this simulations are $\mu_f/\rho= 1.30 \times 10^{-6}$ m$^2$/s, $\phi_m=0.5$, $\phi_b=0.1$, $a=2.82 \, \mu$m, $\alpha=4$, $f(1-\nu)=1.2$, $\beta=1.2$ and $h_0=1.0\times 10^{-12}$ m.}
\end{figure}

Using the same MSBM parameter values we have been using over the last sections, we also compare our simulation results from the MBSM with experimental results reported by Losserand {\it et. al.}~\cite{losserand2019}, who studied lateral migration of red blood cells (RBCs) in confined channel flows. They reported measured RBC velocities in channels of different sizes, and here we compare two of their cases with our simulations, which are shown in figure~\ref{fig:losserand}. Our MSBM approach with the current model parameters predict velocity profiles of different channels sizes comparably well with the experimental velocity profiles.



\begin{figure}[hbt!]
  \centering
  \includegraphics[width=.7\linewidth]{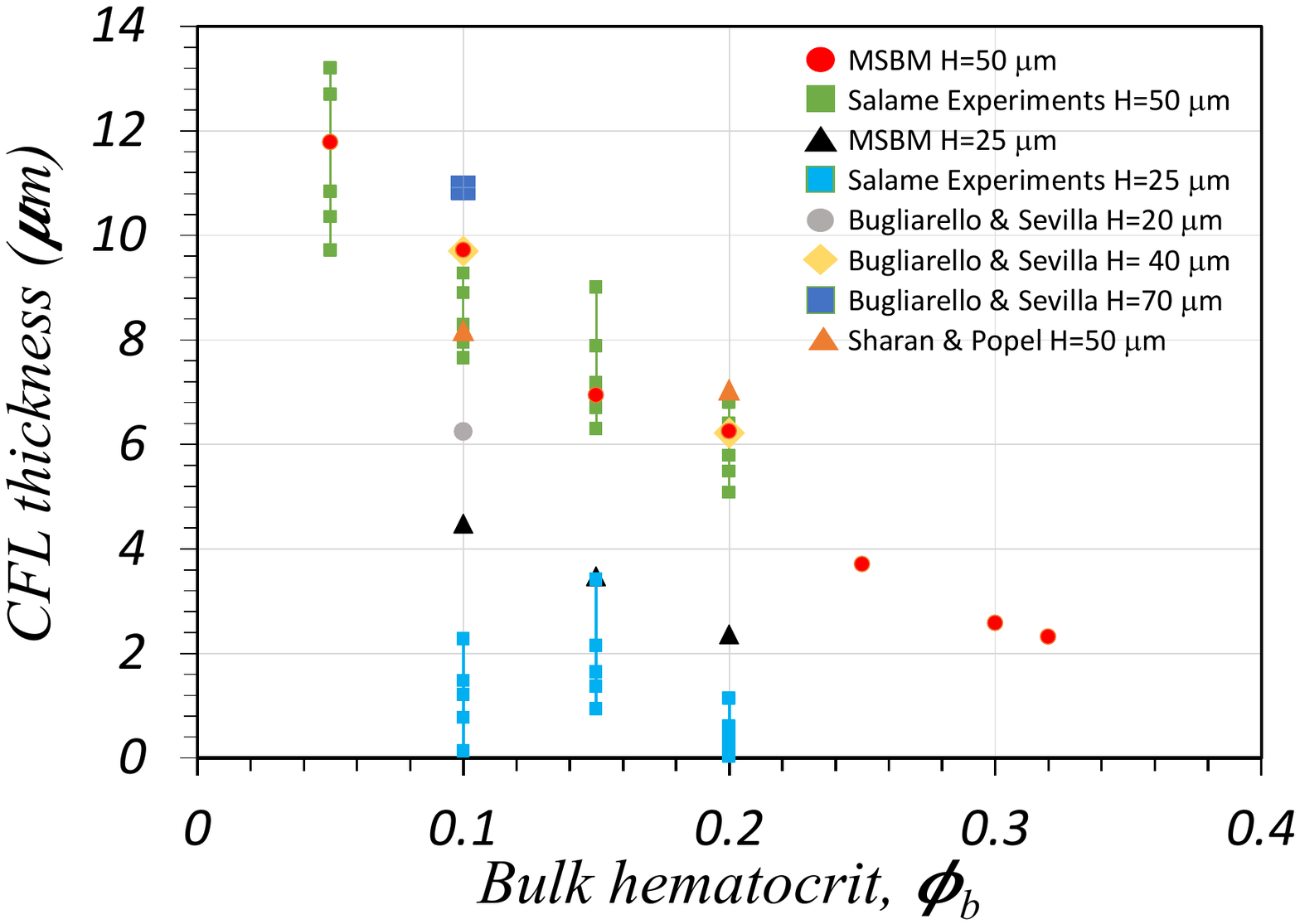}
\caption{\label{fig:CFLvsHemaExpSimA} Comparison between our MSBM simulations and experimental results for the cell-free boundary thickness (CFL) thickness $\delta_{CFL}$ as a function of the bulk hematocrit $\phi_b$. Experimental results reported by Salame~\cite{salame2025} correspond to measurements obtained for RBCs suspended in phosphate-buffered saline
(PBS) in a $50 \, \mu$m channel-height (green squares) and in a $25 \, \mu$m channel-height (blue squares). Other experimental results obtained by Bugliarello \& Sevilla~\cite{bugliarello} are presented as royal blue squares ($H=70 \, \mu$m), yellow diamonds ($H=40 \, \mu$m) and gray circles ($H=20 \, \mu$m). Analytical predictions by Sharan and Popel~\cite{sharan2001} for a $50 \mu$m channel-height are also presented as purple triangles. The MSBM predictions for the larger channel are presented as red circles, while our MSBM results for the smaller channel are shown as black triangles. The MSBM parameter values used are: $\mu_f/\rho= 1.32 \times 10^{-6}$ m$^2$/s, $\beta=1.2$, $\phi_m=0.5$, $\alpha=4$, $a=2.82 \, \mu$m, $f(1-\nu)=1.2$ and $h_0=1.0\times 10^{-12}$ m. }
\end{figure}

To further demonstrate the validity of our method, we compare our CFL data predicted in channel flows with experimental data reported by Salame~\cite{salame2025}. Salame used RBCs suspended in phosphate-buffered saline (PBS) to prevent aggregation, and optical measurements of $\delta_{CFL}$ were obtained through high-speed imaging and image processing, which provided a consistent measurement of the CFL thickness across varying conditions~\cite{salame2025}. 
 
Comparison can be found in figure~\ref{fig:CFLvsHemaExpSimA}, where we plot the experimental CFL data for the idealized RBC suspension (green squares) and our predicted ones. Salame managed to obtain CFL thickness values up to $\phi_b=0.2$. The experimental data covers the distribution of experimental data for each bulk hematocrit: minimum and maximum values, median, lower and upper quartiles. The green squares correspond to the experimental measurements obtained for the channel with a height of $50 \, \mu$m, while the blue squares are the results reported for the $25 \, \mu$m channel height. We also plot the CFL thickness results from the previously shown hematocrit profiles for the $50 \, \mu$m channel (see figure~\ref{fig:hemavsphibvar}), which are shown as red circles. Overall, we can see that the CFL decreases with increasing hematocrit. More importantly, our simulation results are in good agreement with the experimental observations, as our data lie within the distribution range for the Salame data.

In addition, we also compare the MSBM predictions (black triangles) with the experimental measurements reported for the $25 \, \mu$m channel-height (blue squares). Overall, we can observe that these results also follow the trend previously seen for the taller channel, since the CFL thickness tends to decrease as the bulk hematocrit increases. However, the CFL thickness values for the shorter channel are much smaller compared to the $50 \, \mu$m channel-height case. This indicates that the cell-free layer increases with channel height, which is consistent with results reported in the literature~\cite{qi2018time,salame2025}. It can also be seen that with the exception of the case $\phi_b=0.15$, our simulations predicts values of CFL thickness outside the range of values reported by Salame.

In addition, we reported in-vitro experimental data reported by different authors; Sharan \& Popel~\cite{sharan2001} measured the CFL thickness in a 50 $\mu$m tube-height (purple triangles), and their results are also close to our numerical values predicted by our model. Bugliarello \& Sevilla~\cite{bugliarello} carried out experiments using different heights ($H=20,40$ and $70\mu$m) and their results illustrate that the CFL thickness gets thicker as we increase the channel-height, which was also one of the predictions previously obtained from our MSBM simulations.

\subsection{Capturing Physiological Laws in Tubular Flows}
\label{sec:tubularflow}

In this section, we simulate blood flow through a microvessel. A microvessel is a very small blood vessel that is typically found within the microcirculation system, which includes capillaries, arterioles, and venules that are around $10$--$50 \, \mu$m in diameter. These vessels are tubular structures carrying blood through the tissues and organs and are crucial for blood-tissue exchange, delivering oxygen and nutrients while removing waste products.

Our goal then is to use our modified suspension balance model to simulate blood flow through a tube. For simplicity, we limit ourselves to studying 2D axisymmetric flow. For axisymmetric flow, the variables (i.e. velocity, pressure, and volume fraction) do not vary with the angular coordinate $\theta$, and 
we simulate a section of the cylindrical domain (wedge) using a symmetric boundary condition called {\it wedge}.

\begin{figure}[h]
\begin{subfigure}{.5\textwidth}
  \centering
  \includegraphics[width=.95\linewidth]{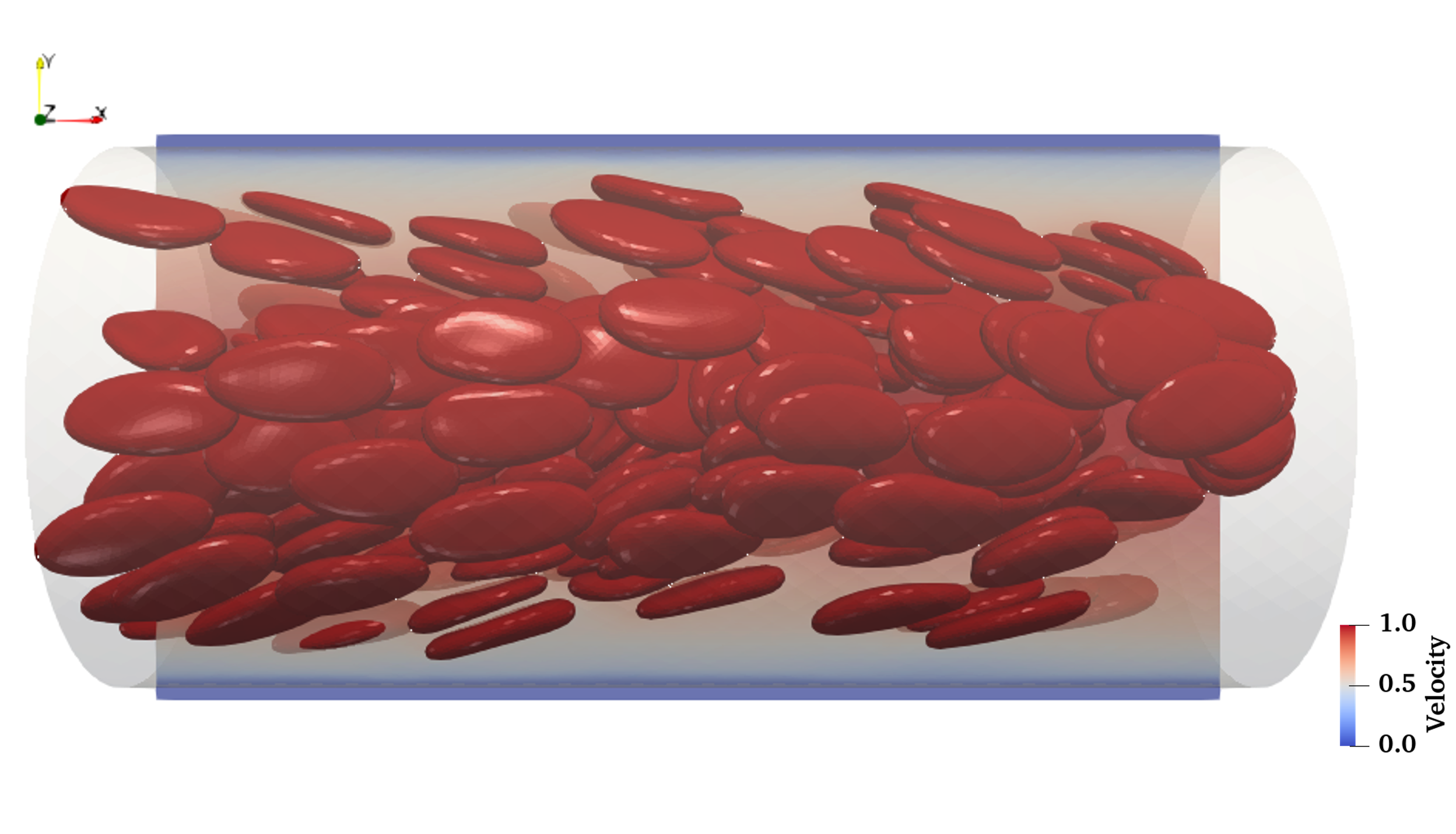}
\end{subfigure}%
\begin{subfigure}{.3\textwidth}
\centering

\tikzset{every picture/.style={line width=0.75pt}} 

\begin{tikzpicture}[x=0.75pt,y=0.75pt,yscale=-1,xscale=1]

\draw  [line width=0.75]  (192.17,77) .. controls (212.73,76.72) and (229.8,95.16) .. (230.3,118.2) .. controls (230.79,141.24) and (214.52,160.16) .. (193.96,160.44) .. controls (173.4,160.73) and (156.33,142.28) .. (155.83,119.24) .. controls (155.34,96.2) and (171.61,77.29) .. (192.17,77) -- cycle ;
\draw   (193.1,120.32) -- (182.52,78.89) -- (202.76,78.74) -- cycle ;
\draw    (193.1,119.63) -- (280.76,81.4) ;
\draw    (202.76,78.74) -- (287.84,43.98) ;
\draw    (182.52,78.89) -- (267.6,44.13) ;
\draw    (267.6,44.13) -- (287.84,43.98) ;
\draw    (287.84,43.98) -- (280.76,81.4) ;
\draw  [dash pattern={on 4.5pt off 4.5pt}]  (267.6,44.13) -- (280.76,81.4) ;
\draw   (224.96,77.67) -- (232.71,74.62) -- (235.98,80.01) -- (228.23,83.06) -- cycle ;
\draw   (245.34,71.23) -- (251.72,68.43) -- (252.7,75.13) -- (246.32,77.93) -- cycle ;
\draw [color={rgb, 255:red, 74; green, 144; blue, 226 }  ,draw opacity=1 ]   (249.02,73.18) .. controls (288.28,86.38) and (285.69,103.03) .. (285.69,108.58) ;
\draw [color={rgb, 255:red, 74; green, 144; blue, 226 }  ,draw opacity=1 ]   (177.14,51.8) .. controls (184.89,62.2) and (211.87,77.93) .. (224.96,77.67) ;
\draw  [dash pattern={on 0.84pt off 2.51pt}]  (280.76,81.4) -- (319.51,61.97) ;
\draw  [draw opacity=0] (188.95,102.31) .. controls (188.54,101.89) and (188.27,101.39) .. (188.19,100.84) .. controls (187.92,99.03) and (189.84,97.36) .. (192.47,97.11) .. controls (195.1,96.86) and (197.45,98.13) .. (197.71,99.94) .. controls (197.83,100.74) and (197.51,101.52) .. (196.89,102.16) -- (192.95,100.39) -- cycle ; \draw   (188.95,102.31) .. controls (188.54,101.89) and (188.27,101.39) .. (188.19,100.84) .. controls (187.92,99.03) and (189.84,97.36) .. (192.47,97.11) .. controls (195.1,96.86) and (197.45,98.13) .. (197.71,99.94) .. controls (197.83,100.74) and (197.51,101.52) .. (196.89,102.16) ;  
\draw [color={rgb, 255:red, 74; green, 144; blue, 226 }  ,draw opacity=1 ]   (142.7,97.58) -- (190.53,104.33) ;
\draw [shift={(193.5,104.75)}, rotate = 188.03] [fill={rgb, 255:red, 74; green, 144; blue, 226 }  ,fill opacity=1 ][line width=0.08]  [draw opacity=0] (5.36,-2.57) -- (0,0) -- (5.36,2.57) -- cycle    ;
\draw    (149.87,142.21) -- (186,122.16) ;
\draw [shift={(188.62,120.71)}, rotate = 150.97] [fill={rgb, 255:red, 0; green, 0; blue, 0 }  ][line width=0.08]  [draw opacity=0] (5.36,-2.57) -- (0,0) -- (5.36,2.57) -- cycle    ;
\draw    (149.87,142.21) -- (149.87,104.28) ;
\draw [shift={(149.87,101.28)}, rotate = 90] [fill={rgb, 255:red, 0; green, 0; blue, 0 }  ][line width=0.08]  [draw opacity=0] (5.36,-2.57) -- (0,0) -- (5.36,2.57) -- cycle    ;
\draw    (149.87,142.21) -- (183.16,151.79) ;
\draw [shift={(186.04,152.62)}, rotate = 196.05] [fill={rgb, 255:red, 0; green, 0; blue, 0 }  ][line width=0.08]  [draw opacity=0] (5.36,-2.57) -- (0,0) -- (5.36,2.57) -- cycle    ;
\draw [color={rgb, 255:red, 74; green, 144; blue, 226 }  ,draw opacity=1 ]   (160.78,79.55) -- (187.8,88.01) ;
\draw [shift={(190.66,88.9)}, rotate = 197.38] [fill={rgb, 255:red, 74; green, 144; blue, 226 }  ,fill opacity=1 ][line width=0.08]  [draw opacity=0] (5.36,-2.57) -- (0,0) -- (5.36,2.57) -- cycle    ;
\draw [color={rgb, 255:red, 74; green, 144; blue, 226 }  ,draw opacity=1 ]   (296.02,36.7) -- (281.69,51.71) ;
\draw [shift={(279.61,53.88)}, rotate = 313.68] [fill={rgb, 255:red, 74; green, 144; blue, 226 }  ,fill opacity=1 ][line width=0.08]  [draw opacity=0] (5.36,-2.57) -- (0,0) -- (5.36,2.57) -- cycle    ;
\draw [color={rgb, 255:red, 208; green, 2; blue, 27 }  ,draw opacity=1 ]   (286.26,78) -- (292.77,47.1) ;
\draw [shift={(293.39,44.17)}, rotate = 101.9] [fill={rgb, 255:red, 208; green, 2; blue, 27 }  ,fill opacity=1 ][line width=0.08]  [draw opacity=0] (5.36,-2.57) -- (0,0) -- (5.36,2.57) -- cycle    ;
\draw [shift={(285.64,80.94)}, rotate = 281.9] [fill={rgb, 255:red, 208; green, 2; blue, 27 }  ,fill opacity=1 ][line width=0.08]  [draw opacity=0] (5.36,-2.57) -- (0,0) -- (5.36,2.57) -- cycle    ;
\draw [color={rgb, 255:red, 208; green, 2; blue, 27 }  ,draw opacity=1 ]   (200.56,121.37) -- (286.04,84.66) ;
\draw [shift={(288.8,83.48)}, rotate = 156.76] [fill={rgb, 255:red, 208; green, 2; blue, 27 }  ,fill opacity=1 ][line width=0.08]  [draw opacity=0] (3.57,-1.72) -- (0,0) -- (3.57,1.72) -- cycle    ;
\draw [shift={(197.81,122.56)}, rotate = 336.76] [fill={rgb, 255:red, 208; green, 2; blue, 27 }  ,fill opacity=1 ][line width=0.08]  [draw opacity=0] (3.57,-1.72) -- (0,0) -- (3.57,1.72) -- cycle    ;
\draw [color={rgb, 255:red, 208; green, 2; blue, 27 }  ,draw opacity=1 ]   (267.69,41.16) -- (284.94,41.16) ;
\draw [shift={(287.94,41.16)}, rotate = 180] [fill={rgb, 255:red, 208; green, 2; blue, 27 }  ,fill opacity=1 ][line width=0.08]  [draw opacity=0] (5.36,-2.57) -- (0,0) -- (5.36,2.57) -- cycle    ;
\draw [shift={(264.69,41.16)}, rotate = 0] [fill={rgb, 255:red, 208; green, 2; blue, 27 }  ,fill opacity=1 ][line width=0.08]  [draw opacity=0] (5.36,-2.57) -- (0,0) -- (5.36,2.57) -- cycle    ;
\draw [color={rgb, 255:red, 74; green, 144; blue, 226 }  ,draw opacity=1 ]   (233.97,38.62) -- (245.21,49.71) ;
\draw [shift={(247.35,51.82)}, rotate = 224.63] [fill={rgb, 255:red, 74; green, 144; blue, 226 }  ,fill opacity=1 ][line width=0.08]  [draw opacity=0] (5.36,-2.57) -- (0,0) -- (5.36,2.57) -- cycle    ;

\draw (257.86,105.78) node [anchor=north west][inner sep=0.75pt]  [font=\fontsize{0.57em}{0.68em}\selectfont] [align=left] {\begin{minipage}[lt]{53.19pt}\setlength\topsep{0pt}
\begin{center}
wedge or symmetry\\(front)
\end{center}

\end{minipage}};
\draw (111.22,29.94) node [anchor=north west][inner sep=0.75pt]  [font=\fontsize{0.57em}{0.68em}\selectfont] [align=left] {\begin{minipage}[lt]{53.19pt}\setlength\topsep{0pt}
\begin{center}
wedge or symmetry\\(back)
\end{center}

\end{minipage}};
\draw (301.17,70.08) node [anchor=north west][inner sep=0.75pt]  [font=\fontsize{0.57em}{0.68em}\selectfont] [align=left] {Axis of symmetry};
\draw (106.63,108.45) node [anchor=north west][inner sep=0.75pt]  [font=\fontsize{0.56em}{0.67em}\selectfont]  {$< 5 ^{\circ}$};
\draw (186.96,144.44) node [anchor=north west][inner sep=0.75pt]  [font=\fontsize{0.53em}{0.64em}\selectfont]  {$z$};
\draw (137.4,101.75) node [anchor=north west][inner sep=0.75pt]  [font=\fontsize{0.53em}{0.64em}\selectfont]  {$y$};
\draw (186.62,122.71) node [anchor=north west][inner sep=0.75pt]  [font=\fontsize{0.5em}{0.6em}\selectfont]  {$x$};
\draw (137.93,68.32) node [anchor=north west][inner sep=0.75pt]  [font=\fontsize{0.57em}{0.68em}\selectfont] [align=left] {inlet};
\draw (74.93,63.5) node [anchor=north west][inner sep=0.75pt]  [font=\tiny]  {$\phi ( x=0) =\phi _{b}$};
\draw (197.23,20.22) node [anchor=north west][inner sep=0.75pt]  [font=\tiny]  {$u_{x}( H) \ =0$};
\draw (202.89,10.82) node [anchor=north west][inner sep=0.75pt]  [font=\tiny]  {$\partial \phi /\partial y=0$};
\draw (204.98,-1.31) node [anchor=north west][inner sep=0.75pt]  [font=\tiny]  {$\mathbf{j}_{\perp }^{\prime } \cdotp \mathbf{n} =0$};
\draw (217.13,28.44) node [anchor=north west][inner sep=0.75pt]  [font=\fontsize{0.57em}{0.68em}\selectfont] [align=left] {wall};
\draw (289.75,24.66) node [anchor=north west][inner sep=0.75pt]  [font=\fontsize{0.57em}{0.68em}\selectfont] [align=left] {outlet};
\draw (315.73,16.6) node [anchor=north west][inner sep=0.75pt]  [font=\tiny]  {$\partial \phi /\partial x=0$};
\draw (317.72,33.88) node [anchor=north west][inner sep=0.75pt]  [font=\tiny]  {$\partial u_{z} /\partial x=0$};
\draw (324.91,25.39) node [anchor=north west][inner sep=0.75pt]  [font=\tiny]  {$p=0$};
\draw (70.26,71.93) node [anchor=north west][inner sep=0.75pt]  [font=\tiny]  {$u_{x}( x=0) =\mathrm{U_{0}}$};
\draw (289.7,51.66) node [anchor=north west][inner sep=0.75pt]  [font=\fontsize{0.59em}{0.71em}\selectfont]  {$R$};
\draw (243.3,103.02) node [anchor=north west][inner sep=0.75pt]  [font=\fontsize{0.59em}{0.71em}\selectfont]  {$L$};
\draw (267.38,25.88) node [anchor=north west][inner sep=0.75pt]  [font=\fontsize{0.59em}{0.71em}\selectfont]  {$W$};
\draw (88.56,85.89) node [anchor=north west][inner sep=0.75pt]  [font=\fontsize{0.57em}{0.68em}\selectfont] [align=left] {\begin{minipage}[lt]{35.53pt}\setlength\topsep{0pt}
\begin{center}
wedge angle
\end{center}

\end{minipage}};

\end{tikzpicture}

\end{subfigure}
\caption{\label{fig:tubegeom} Tubular flow set-up. The snapshot (left) that illustrates the RBCs in the tubular flow was obtained using our in-house multiscale blood flow solver (described in section~\ref{sec:multiscalesol}) in a $40 \, \mu$m diameter tube with bulk hematocrit $\phi_b=0.2$. The schematic (right) illustrates the flow geometry and the boundary conditions imposed in our {\it OpenFOAM} solvers.}
\end{figure}
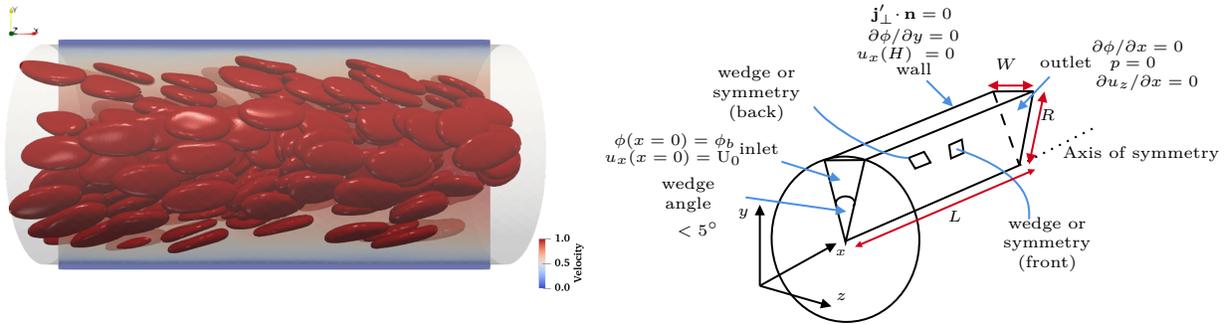


The cross-section of the tube used in our simulations is shown in figure~\ref{fig:tubegeom}: for our 2D axisymmetric cylinder, the geometry is specified as a wedge of small angle and and 1 cell thick (width $W$), running along the centerline, straddling one of the coordinate planes. This wedge has a length $L$ and a radius $R$, and the boundary conditions are similar to the ones specified for channel flows: at the walls ($y=R$), we have no slip boundary conditions for the velocity field, zero-gradient boundary conditions for both the pressure and volume fraction, and we impose a $\textit{slip}$ boundary condition for the vector $\bm{j}_{\perp}^{\, \prime}$. For initial conditions in the transient solution scheme, the pressure, velocity vector, and stress tensor are all set to zero. The inlet boundary conditions at both the inlet and outlet depends on the set up we use to solve the governing equations, which is discussed below.


Since our global $Re$ and particle $Re_p$ Reynolds numbers values for these simulations lie in the following ranges: $Re=0.038-0.098$ and $Re_p=1.44 \times 10^{-4}-3.68 \times 10^{-4}$, we neglect inertia and we consider the effect of the walls on the blood flow (see equation~(\ref{eq:lift_force12})) and we ensure that the aspect ratio $L/R$ is large enough to guarantee that our profiles will be fully developed at the outflow (see equation~(\ref{eq:aspectratio})). 
We simulate blood flow using the MSBM in a tube with a diameter of $40 \, \mu$m using the following parameter values: $\mu_f/\rho= 1.30 \times 10^{-6}$ m$^2$/s, $\phi_m=0.5$, $\beta=1.2$, $\alpha=4$, $a=2.82 \, \mu$m, $f(1-\nu)=1.2$ and $h_0=1.0\times 10^{-12}$ m. A inlet velocity value of $U_0=0.005$ m/s was used in all of our simulations, resulting in wall shear rate values of $\gd_w \approx 1000-1600$ s$^{-1}$. This range of shear rate is commonly observed in arterioles and microvessels~\cite{kroll1996,liu2018nanoparticle}. 

\begin{figure}[hbt!]
\begin{subfigure}{.5\textwidth}
  \centering
  \includegraphics[width=.9\linewidth]{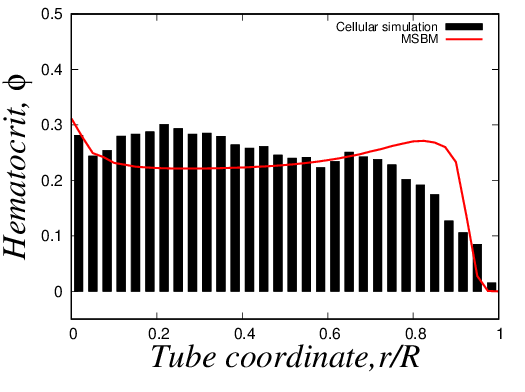}
  \caption{$t=0.008$ s}
\end{subfigure}%
\begin{subfigure}{.5\textwidth}
  \centering
  \includegraphics[width=.9\linewidth]{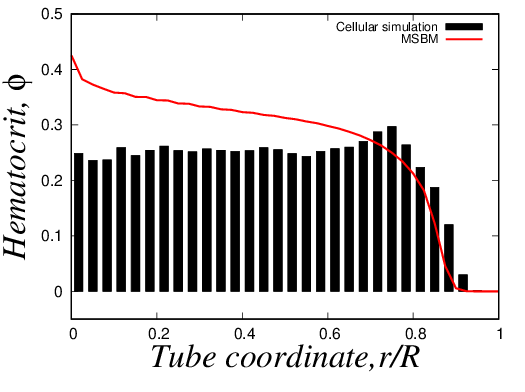}
  \caption{$t=0.11$ s}
\end{subfigure}
\begin{subfigure}{.5\textwidth}
  \centering
  \includegraphics[width=.9\linewidth]{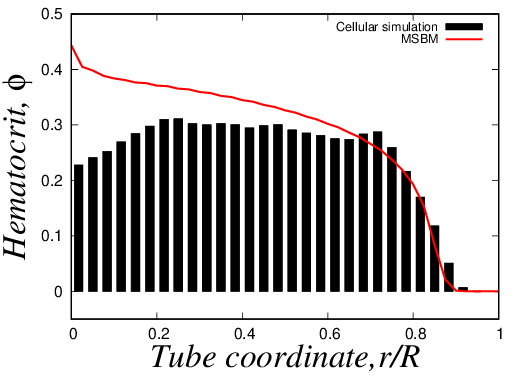}
  \caption{$t=0.27$ s}
\end{subfigure}%
\begin{subfigure}{.5\textwidth}
  \centering
  \includegraphics[width=.9\linewidth]{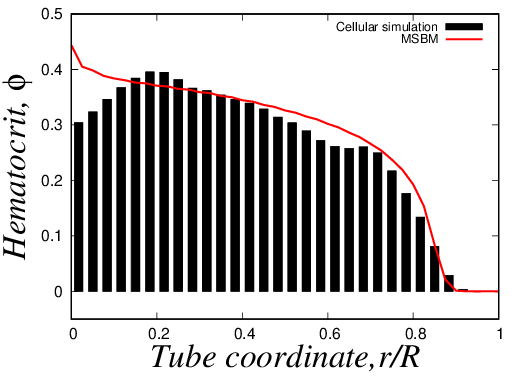}
  \caption{$t=0.53$ s}
\end{subfigure}
\begin{subfigure}{.5\textwidth}
  \centering
  \includegraphics[width=.9\linewidth]{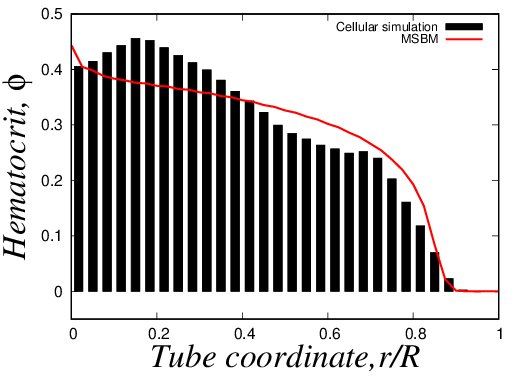}
  \caption{Steady-state}
\end{subfigure}%
\begin{subfigure}{.5\textwidth}
  \centering
  \includegraphics[width=.9\linewidth]{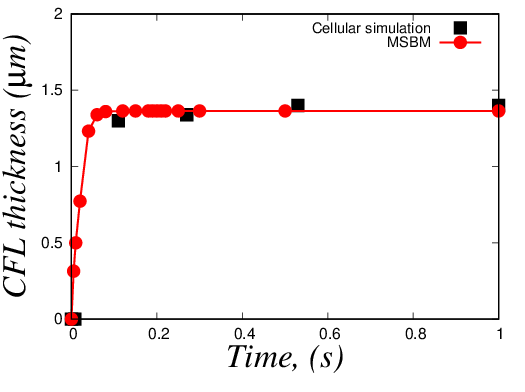}
  \caption{CFL vs time}
  \label{fig:cflvstime}
\end{subfigure}
\caption{\label{fig:msbmdisctemp} Comparison between the continuum model (MSBM, red lines) and our discrete simulation (black bars) results for the temporal evolution of the hematocrit in a $40 \, \mu$m diameter tube with an bulk hematocrit value of $\phi_b = 0.2$. These snapshots were taken at different $t$. The bottom right figure illustrates the evolution in time of the CFL thickness predicted by both the MSBM and the cellular blood flow simulation. The MSBM parameter values used are $\mu_f/\rho= 1.30 \times 10^{-6}$ m$^2$/s, $\phi_m=0.50$, $\beta=1.20$, $\alpha=4$, $a=2.82 \, \mu$m, $f(1-\nu)=1.2$ and $h_0=1.0\times 10^{-12}$ m. }
\end{figure}

To illustrate that the MSBM is able to capture transient microscale hemodynamics in tubular flows, we compare our transient MSBM simulations with those from the discrete RBCs simulations. As previously described, in the setup used by our multiscale blood flow solver, periodic boundary conditions are used, so that if the RBCs exit the tube, they immediately re-enter from the opposite side (inlet), and thus, the tube hematocrit always remains constant. To properly compare our MSBM simulations with our discrete RBCs simulations, we impose the same cyclic boundary conditions at both the inlet and the outlet for all the flow quantities (velocity, pressure, volume fraction, and flux of mass). To drive the flow, we fix a velocity value of $0.005$ m/s in the 'patchMeanVelocityForce’ specified in the fvConstraints file, which effectively adds a velocity-matching pressure gradient or body force as a source term in the momentum equation. A bulk hematocrit value of $\phi_b=0.2$ is imposed in the whole channel. The average tube hematocrit $H_t$ (computed as $H_t =\int_{0}^{R} \big[2\,\pi \, r \, \phi (r) \big]  \,dr/\pi \, r^2$) is calculated to have the same value as $\phi_b=0.2$, confirming the effect of the cyclic/periodic boundary condition. Our results can be found in figure~\ref{fig:msbmdisctemp}.
 



The discrete simulation results are shown as black bars while the MSBM results are shown as red lines. Similarly to the channel-flow case (see section~\ref{sec:temp}), overall, we observe good agreement between the results predicted by the mesoscale and continuum model, especially near the wall where both predict roughly the same CFL thickness steady-state value: the continuum simulation predicts a value of $\delta_{CFL}\approx 1.36 \mu$m, while the discrete simulation gives a value of $\delta_{CFL}\approx 1.41 \mu$m. At the beginning of the simulations ($t=0.008$ s), we see that both the continuum and the discrete results predict a semi-uniform hematocrit near the centerline, and while and the discrete simulations show that the flow is not yet cell-depleted near the tube walls, the MSBM simulations predict a very thin CFL. At $t=0.11$ s, we start to observe the formation of the CFL for both cases, though the hematocrit seems to remain fairly constant in the discrete simulations, while the MSBM starts to predict more migration of RBCs towards the centerline. At times $t=0.27\sim0.53$ s, we observe a overall good match in our results across the tube except near the centerline, where the cellular blood flow simulations still lag behind in catching up the hematocrit level predicted by the MSBM. 

Once a steady state has been reached, we see that both methodologies predict roughly the same value of hematocrit at the centerline, but the maximum hematocrit value predicted by the discrete simulations is not seen at the centerline. The decreasing hematocrit towards the centerline, which is not seen in channel flows (see figure~\ref{fig:msbmdisctemp}), could be due to the finite size effect of RBCs in the discrete simulation prevents the RBCs from occupying the centerline space. 

The temporal evolution of the CFL thickness predicted by the MSBM is also illustrated in figure~\ref{fig:cflvstime} (see red solid line), where it can be seen that a cell-depleted region forms since the beginning of the flow and increases its size very fast when $t < 0.1$ s, and then reaches equilibrium around $t \approx 0.2$ s. The CFL evolution predicted by MSBM quantitatively match well with the cellular blood simulations (black solid symbols).


\begin{figure}[hbt!]
\begin{subfigure}{.5\textwidth}
  \centering
  \includegraphics[width=.9\linewidth]{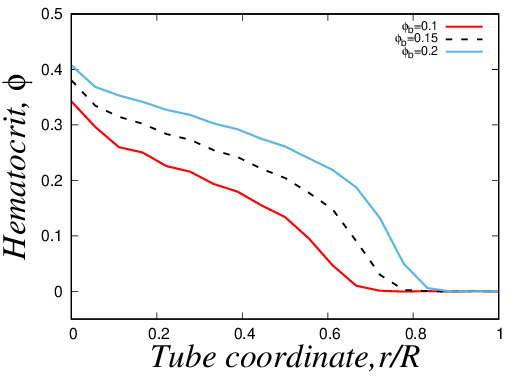}
  \caption{$\phi$ vs $r/R$}
  \label{fig:hemavsrtube}
\end{subfigure}%
\begin{subfigure}{.5\textwidth}
  \centering
  \includegraphics[width=.9\linewidth]{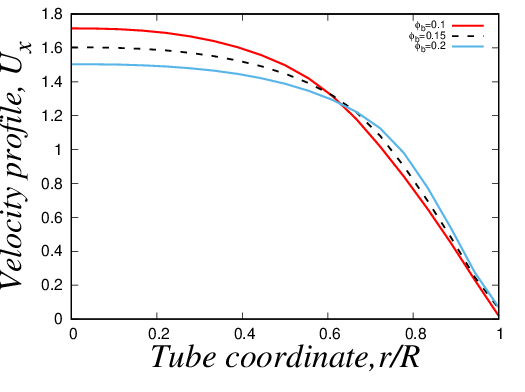}
  \caption{$u_z$ vs $r/R$}
  \label{fig:UxvsRTUBE}
\end{subfigure}%
\caption{\label{fig:tubularprofiles} Hematocrit (left) and velocity (right) profiles obtained using our $\textit{SbmLiftFoam}$ solver for different values of bulk hematocrit $\phi_b$ in tubular flow. These curves were obtained in a tube with a $20 \, \mu$m radius. The blood parameters used in our simulations are: $\mu_f/\rho= 1.30 \times 10^{-6}$ m$^2$/s, $\phi_m=0.5$, $\beta=1.20$, $\alpha=4$, $a=2.82 \, \mu$m, $f(1-\nu)=1.2$ and $h_0=1.0\times 10^{-12}$ m. From top to bottom: $\phi_b=0.20, 0.15, 0.1$.}
\end{figure}

We also study the effect of variable bulk hematocrit $\phi_b$ in tubular flows, where our results are reported in figure~\ref{fig:tubularprofiles}. To obtain these results, we use our standard approach: at the inlet, we impose a uniform velocity profile $U_0$, for the volume fraction, we impose the value of the hematocrit ($\phi_b$), and zero gradient boundary conditions for both the pressure and the flux vector $\bm{j}_{\perp}^{\, \prime}$. Finally, at the outlet, we set a pressure value equal to zero and fully developed boundary conditions for the other flow quantities. For initial conditions in the transient solution scheme, the pressure, velocity vector, and stress tensor are all set to zero. The volume fraction along the tube will have the value of the bulk hematocrit $\phi_b$ imposed at the inlet.  

In figure~\ref{fig:tubularprofiles}, we show the hematocrit and velocity profiles for different values of bulk volume fraction. Our mesh has 18 cells in the radial direction, 1 cell to create the wedge, and 100 cells in the flow direction. 
Similarly to the channel-flow case, we observe migration of the RBCs towards the core region of the tube (see figure~\ref{fig:hemavsrtube}), and higher values of volume fraction at this location will be reached when the bulk volume fraction is higher. For all the three cases of bulk hematocrit, the MSBM predicts the formation of a CFL, and its thickness will increase if we decrease the bulk volume fraction, which is consistent with our previously obtained results in the channel geometry shown in section~\ref{sec:hemavar}.  
For the velocity profiles (see figure~\ref{fig:UxvsRTUBE}), we also observe a plug flow region near the centerline for the case with greater values of $\phi_b$, while the velocity will tend to adopt a more parabolic shape if we decrease the bulk concentration.

\subsubsection{F\aa hr\ae us Effect}
\label{sec:fahraeuseff}

To test the physiological relevance of our model, we further tested our prediction against well-known hemorheological principles such as the {\it F\aa hr\ae us effect}, which describes the discharge hematocrit exceeding the (average) tube hematocrit due to RBC-laden region discharges faster than the average outflow plasma.
Quantitatively, Pries {\it et al.}~\cite{pries1990} developed an empirical correlation to quantify the ratio of the tube hematocrit $H_t$ versus the discharge hematocrit $H_d$ given the tube diameter $D$ (in $\mu$m) and discharged hematocrit are known, which read,

\begin{equation}
\frac{H_t}{H_d} = H_d +(1-H_d)\big[1+1.7 \exp(-0.415\,D) -0.6\exp(-0.011\,D)\big].
\label{eq:prieseqn}
\end{equation}

\begin{figure}[hbt!]
  \centering
  \includegraphics[width=.6\linewidth]{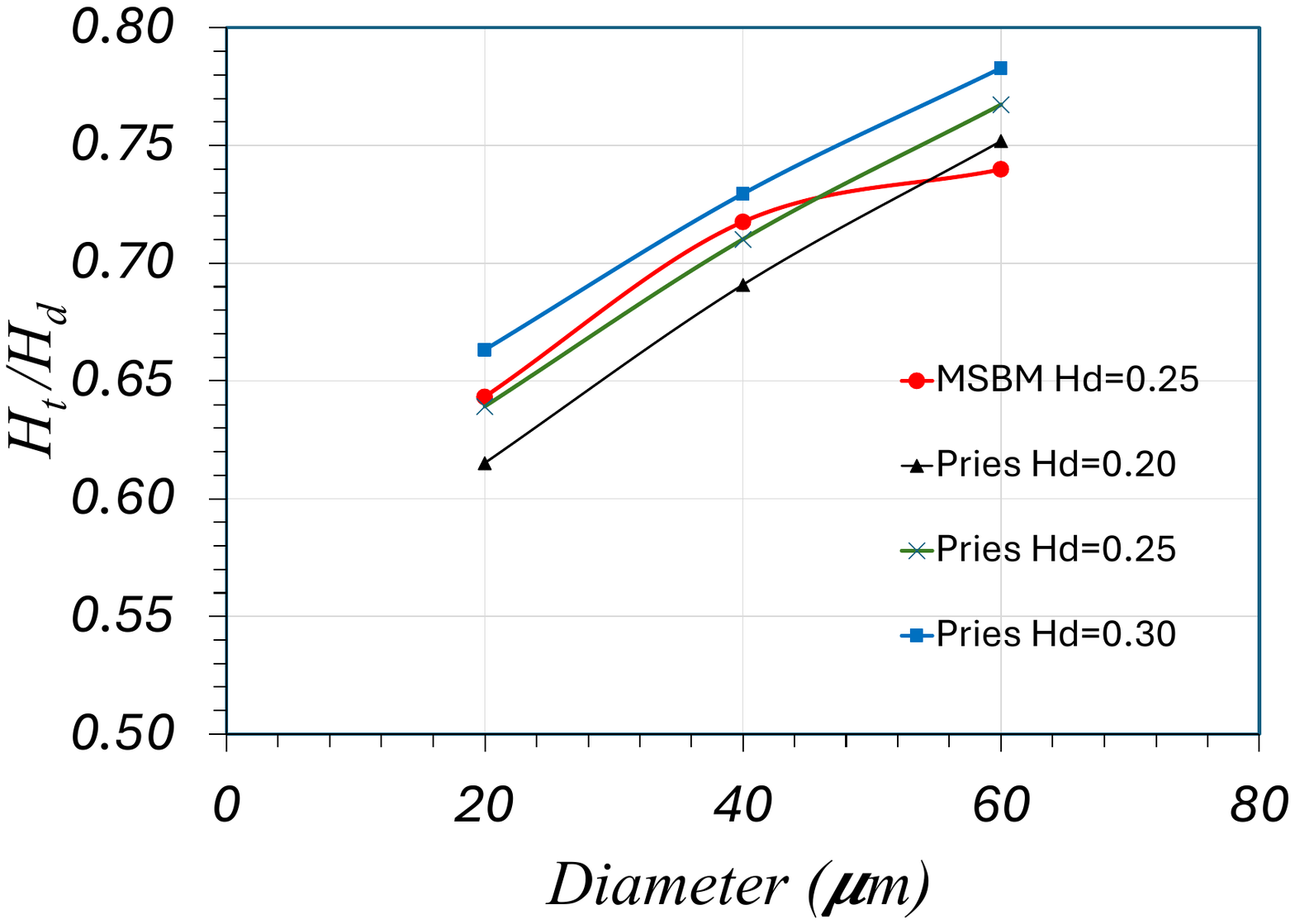}
\caption{\label{fig:fahraeuseff} Ratio of tube hematocrit to discharge hematocrit $H_t/H_d$ as a function of the tube diameter $D$. The red circles represent the predictions from the MSBM using a discharge hematocrit value of $H_d=0.25$. The rest of the lines represent the theoretical predictions from Pries' expression (see equation~(\ref{eq:prieseqn}): blue squares ($H_d=0.30$), green crosses ($H_d=0.24$), black triangles ($H_d=0.20$).}
\end{figure}

As discussed in \S \ref{sec:temp}, by imposing a fixed hematocrit value $\phi_b$ at the inlet and a zero-gradient boundary condition at the outlet, we can capture the decrease or the discharge of the tube average hematocrit (computed as $H_t =\int_{0}^{R} \big[2\,\pi \, r \, \phi (r) \big]  \,dr/\pi \, r^2$) at the outlet, as described by the F\aa hr\ae us Effect. The discharged hematocrit at the outlet can be calculated as the velocity-weighted average hematocrit, which is computed as $H_d =\int_{0}^{R} \big[2\,\pi \, r \,\phi (r) \, U_x(r) \big]  \,dr/\int_{0}^{R} \big[2\,\pi \, r \, U_x(r) \big] \,dr$ and remains equal to $\phi_b$ according to mass conservation.

We carry out MSBM simulations using different tube diameters ($D=20,40$ and $60 \, \mu$m, and the following model parameter values are used: $\mu_f/\rho= 1.30 \times 10^{-6}$ m$^2$/s, $\phi_m=0.50$, $\phi_b=0.25$, $\beta=1.2$, $\alpha=4$, $a=2.82 \, \mu$m, $f(1-\nu)=1.2$ and $h_0=1.0\times 10^{-12}$ m. Once steady-state is reached, we obtain the hematocrit profile $\phi(r)$ to compute the tube hematocrit $H_t$, which is then used to calculate the ratio $H_t/H_d$. These results from our simulations are compared with the predictions from Pries' expression (see equation~(\ref{eq:prieseqn})), which can be found in figure~\ref{fig:fahraeuseff}. 
For our MSBM simulations, we used a fixed value of discharge hematocrit of $H_d=0.25$ (red circles), and we obtained different curves with three different values of discharge hematocrit using Pries' equation: $H_d=0.2$ (black triangles), $H_d=0.25$ (green crosses) and $H_d=0.3$ (blue squares). We notice that for tubes with a diameter smaller than $60 \, \mu$m, our MSBM results can well capture both the increasing trend as well as the magnitude of the ratio $H_t/H_d$ compared to those predicted by Pries' correlation. However, this tendency tends to be weakening for tubes larger than $40 \, \mu$m.


\begin{figure}[hbt!]
  \centering
  \includegraphics[width=.6\linewidth]{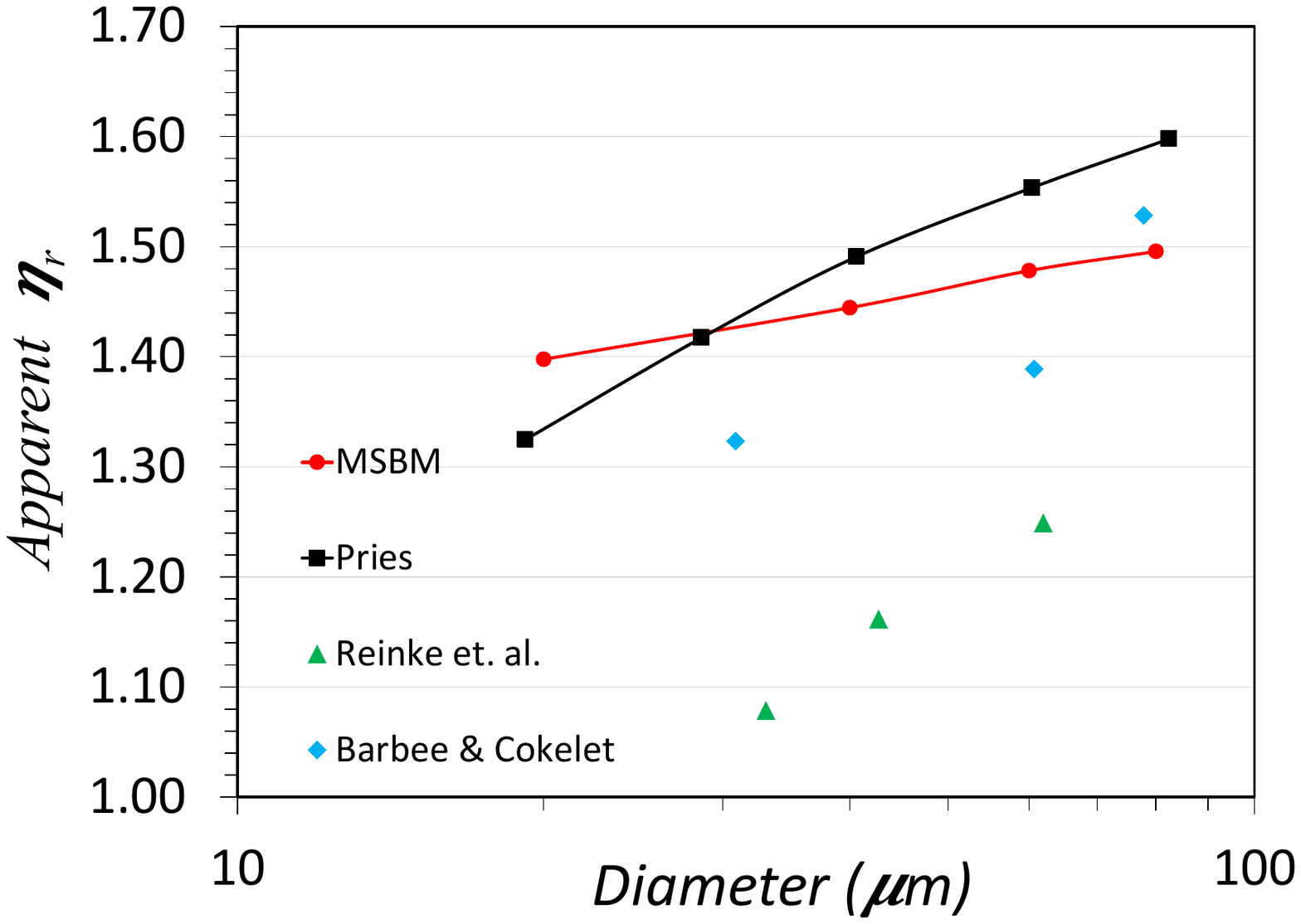}
\caption{\label{fig:fahraeuseLindff} Apparent relative viscosity of blood $ \eta_r$ as a function of the tube diameter $D$. The red circles represent the predictions from the MSBM, the black squares represent the experiments from Pries~\cite{secomb2013}, the blue diamonds are the experimental results of Barbee \& Cokelet~\cite{barbee197} and the green triangles correspond to Reinke et. al. results~\cite{reinke1986}.}
\end{figure}

\subsubsection{F\aa hr\ae us-Linqvist Effect}
\label{sec:fahraeuslinqvisteff}
The Fåhræus–Lindqvist effect describes how the apparent viscosity of blood decreases when it flows through smaller-diameter vessels, particularly those between about 10 µm and 300 µm in diameter (the scale of arterioles and capillaries)~\cite{secomb2013}. To test the physiological relevance of our model, we also study the blood viscosity dependence on the tube diameter. We carried out simulations for different radii $R=D/2$ using the same MSBM parameters described above. Once steady state is reached, we obtain the pressure gradient $\Delta p/L$ and the volumetric flow $Q=\int_{0}^{R} \big[ 2 \pi r \, U_x(r) \big]  \,dr$ to compute the blood apparent viscosity $\eta_b$ using the Hagen-Poiseuille equation:

\begin{equation}
\eta_b=\frac{\Delta p \,\pi\,R^4 }{8\,L\,Q} \qquad \eta_r= \frac{\eta_b}{\eta_f},
\end{equation}
Once $\eta_b$ is obtained, we can compute the apparent relative viscosity of blood $ \eta_r$, which is defined as the ratio between $\eta_b$ and the plasma viscosity $\eta_0$.
We compare our predicted apparent viscosity with various sources of experimental results~\cite{secomb2013,barbee197,reinke1986} shown in figure~\ref{fig:fahraeuseLindff}. It can be shown that our continuum model results overall agree well with the experimental observations due to the well capture of the cell free layer. The slightly weaker dependence of the apparent viscosity and the discharge hematocrit ratio (in the previous section) on the change of diameter may be due to the lack of particle-screen effect in the lift force model, which can improved by incorporating distance-dependent lift-distance scalings~\cite{bureau2023lift}.



\section{Concluding remarks} \label{sec:conclusions}

We demonstrate that the conventional suspension balance model can be modified to model deformable particle suspension physics under confinement by incorporating a lift force to account for the deformability-induced hydrodynamic force pushing the particle away from the wall. The modified SBM was applied to simulate physiological blood flows under various flow configurations including simple shear flow, straight channel and tube flows. Our model captures the plug-flow velocity profiles as well as the cell-free layer phenomena as a key characteristic of non-Newtonian hemorheology bahaviors, also quantitatively matching those reported in experiments.  
In addition to the equilibrium CFL prediction, the MSBM can also capture the transient behavior of RBC migration that matches well with experimentally validated discrete RBC suspension simulations. Under simple shear flow, at the short-time scales when RBC-RBC interaction remains weak, we find similar universal linear scale of the cubic CFL thickness with respect to time, comparable to the scaling found with a single RBC migration experiment by Grandchamp {\it et. al.}~\cite{grandchamp2013}.

The MSBM model generally captures the trend of the F\aa hr\ae us and the F\aa hr\ae us-Linqvist effects in tubular with reasonable agreement. However, beyond the tube size of 40 $\mu$m, there appears to be an overprediction of the CFL, which we conjecture to be the result of the lack of the physical screening effect in the RBC-laden region that overestimates the lift force as RBCs migrate away from the wall. Imposing a distance-dependence of the lift force exponent as discussed recently in Bureau et al.~\cite{bureau2023lift} may be able resolve this discrepancy. We also argue that the lack of this distance-dependent screen effect could potentially explain the absence of the secondary hematocrit peaks in the MSBM results near the wall, contrary to what is observed in the cellular scale simulation. Notably, Yeo and Martin~\cite{yeo2010,yeo2011} previously attributed the secondary peak near the wall to the strong particle-wall lubrication interaction being locally present near the wall while absent in the particle laden region, in alignment with the conjecture of the local particle screen effect. Incorporating this particle screening effect with particle-wall distance dependence in the lift force is a subject of our future work.

In sum, our current work demonstrates that the lift-force augmented suspension balance model can well capture the non-Newtonian hemorheology under large confinement and has the potential to further advance efficient, at-scale biomedical and industrial applications.

\section*{Acknowledgments}
The authors acknowledge financial support from the Sandia National Laboratories LDRD-SUPN award, FAMU-FSU Engineering Start-Up Fund, and the computational resources provided by the Florida State University Research Computing Center. Sandia National Laboratories is a multimission laboratory managed and operated by National Technology and Engineering Solutions of Sandia LLC, a wholly owned subsidiary of Honeywell International Inc. for the U.S. Department of Energy’s National Nuclear Security Administration
under contract DE-NA0003525. This paper describes objective technical results and analysis. Any subjective views or opinions that might be expressed in the paper do not necessarily represent the views of the U.S. Department of Energy or the United States Government.

\section*{Contribution}
Conceptualization: RRR, ZLL\\
Data curation: HACS, RT\\
Formal analysis: HACS, RRR, ZLL\\
Funding acquisition: RRR, ZLL\\
Investigation: HACS, RRR, ZLL\\
Methodology: HACS, WO, RM, RRR, ZLL\\
Project administration: RRR, ZLL\\
Resources: RRR, ZLL\\
Supervision: RRR, ZLL\\
Validation: HACS, WO, RM, RRR, ZLL\\
Visualization: HACS, RT\\
Writing – original draft: HACS, WO, RM, RR, ZLL\\
Writing – review \& editing: HACS, RR, ZLL\\
\bibliographystyle{unsrt}
\bibliography{bibliography}

\end{document}